\input harvmac.tex

\vskip 1.5in
\Title{\vbox{\baselineskip12pt
\hbox to \hsize{\hfill}
\hbox to \hsize{\hfill WITS-CTP-041}}}
{\vbox{
	\centerline{\hbox{Ground Ring of $\alpha$-Symmetries
		}}\vskip 5pt
        \centerline{\hbox{and Sequence of RNS String Theories
		}} } }
\centerline{Dimitri Polyakov\footnote{$^\dagger$}
{dimitri.polyakov@wits.ac.za}}
\medskip
\centerline{\it National Institute for Theoretical Physics (NITHeP)}
\centerline{\it  and School of Physics}
\centerline{\it University of the Witwatersrand}
\centerline{\it WITS 2050 Johannesburg, South Africa}
\vskip .3in

\centerline {\bf Abstract}
We construct a sequence of nilpotent BRST charges
in RNS superstring theory, based on new gauge symmetries
on the worldsheet, found in this paper.
 These new local gauge symmetries originate from the
 global nonlinear space-time $\alpha$-symmetries,
 shown to form a
noncommutative ground ring in this work.
The important subalgebra of these symmetries is $U(3)\times{X_6}$,
where $X_6$ is solvable Lie algebra consisting of 6 elements
with commutators reminiscent of the Virasoro type.
We argue that the new BRST charges found in this work
describe the kinetic terms in string field
theories around curved backgrounds of the $AdS\times{{CP}_n}$-type,
determined by the geometries of hidden extra dimensions induced
by the global $\alpha$-generators.
The identification of these backgrounds is however left for the work
in progress. 
\Date{May 2009}
\vfill\eject
\lref\berk{N. Berkovits, JHEP 0801:065(2008)}
\lref\nberk{N. Berkovits, JHEP 0004:018 (2000)}
\lref\nberkk{ N. Berkovits, Phys. Lett. B457 (1999)}
\lref\nberkkk{N.Berkovits, JHEP 0108:026 (2001)}
\lref\green{M.B. Green, J.H. Schwarz, Phys. Lett. B136 (1984) 367}
\lref\fms{D. Friedan, E. Martinec, S. Shenker, Nucl. Phys. B271 (1986) 93}
\lref\sieg{W. Siegel, Nucl. Phys. B263 (1986) 93}
\lref\howe{ P. Howe, Phys. Lett. B258 (1991) 141}
\lref\phowe{ P. Howe, Phys. Lett. B273 (1991) 90}
\lref\verl{H. Verlinde, Phys.Lett. B192:95(1987)}
\lref\bars{I. Bars, Phys. Rev. D59:045019(1999)}
\lref\barss{I. Bars, C. Deliduman, D. Minic, Phys.Rev.D59:125004(1999)}
\lref\barsss{I. Bars, C. Deliduman, D. Minic, Phys.Lett.B457:275-284(1999)}
\lref\lian{B. Lian, G. Zuckerman, Phys.Lett. B254 (1991) 417}
\lref\pol{I. Klebanov, A. M. Polyakov, Mod.Phys.Lett.A6:3273-3281}
\lref\wit{E. Witten, Nucl.Phys.B373:187-213  (1992)}
\lref\grig{M. Grigorescu, math-ph/0007033, Stud. Cercetari Fiz.36:3 (1984)}
\lref\witten{E. Witten,hep-th/0312171, Commun. Math. Phys.252:189  (2004)}
\lref\wb{N. Berkovits, E. Witten, hep-th/0406051, JHEP 0408:009 (2004)}
\lref\zam{A. Zamolodchikov and Al. Zamolodchikov,
Nucl.Phys.B477, 577 (1996)}
\lref\mars{J. Marsden, A. Weinstein, Physica 7D (1983) 305-323}
\lref\arnold{V. I. Arnold,''Geometrie Differentielle des Groupes de Lie'',
Ann. Inst. Fourier, Grenoble 16, 1 (1966),319-361}
\lref\self{D. Polyakov,  Int.J.Mod. Phys A20: 2603-2624 (2005)}
\lref\selff{D. Polyakov, Phys. Rev. D65: 084041 (2002)}
\lref\ampf{S.Gubser,I.Klebanov, A.M.Polyakov,
{ Phys.Lett.B428:105-114}}
\lref\malda{J.Maldacena, Adv.Theor.Math.Phys.2 (1998)
231-252, hep-th/9711200} 
\lref\sellf{D. Polyakov, Int. J. Mod. Phys A20:4001-4020 (2005)}
\lref\selfian{I.I. Kogan, D. Polyakov, Int.J.Mod.PhysA18:1827(2003)}
\lref\doug{M.Douglas et.al. , hep-th/0307195}
\lref\dorn{H. Dorn, H. J. Otto, Nucl. Phys. B429,375 (1994)}
\lref\prakash{J. S. Prakash, 
H. S. Sharatchandra, J.Math.Phys.37:6530-6569 (1996)}
\lref\dress{I. R. Klebanov, I. I. Kogan, A. M.Polyakov,
Phys. Rev. Lett.71:3243-3246 (1993)}
\lref\selfdisc{ D. Polyakov, hep-th/0602209, to appear
in IJMPA}
\lref\wittwist{E. Witten, Comm. Math.Phys.252:189-258 (2004)}
\lref\wittberk{ N. Berkovits, E. Witten, JHEP 0408:009 (2004)}
\lref\cachazo{ F. Cachazo, P. Svrcek, E. Witten, JHEP 0410:074 (2004)}
\lref\barstwist{ I. Bars, M. Picon, Phys.Rev.D73:064033 (2006)}
\lref\barstwistt{ I. Bars, Phys. Rev. D70:104022 (2004)}
\lref\selftwist{ D. Polyakov, Phys.Lett.B611:173 (2005)}
\lref\klebwitt{I. Klebanov, E.Witten, Nucl.Phys.B 556 (1999) 89}
\lref\klebgauge{C. Herzog, I. Klebanov, P. Ouyang, hep-th/0205100}
\lref\ampconf{A. M. Polyakov, Nucl.Phys.B486(1997) 23-33}
\lref\amplib{A. M. Polyakov, hep-th/0407209,in 't Hooft, G. (ed.):
50 years of Yang-Mills theory 311-329}
\lref\wit{E. Witten{ Adv.Theor.Math.Phys.2:253-291,1998}}
\lref\alfa{D. Polyakov, Int.J.Mod.Phys.A22:5301-5323(2007)}
\lref\ghost{D. Polyakov, Int.J.Mod.Phys.A22:2441(2007)}
\lref\selfgauge{D. Polyakov, Int.J.Mod.Phys. A24:113(2009)}
\lref\abjm{O. Aharony, O. Bergman, D. Jafferis, J. Maldacena,
JHEP 0810:091 (2008)}
\lref\bagger{ J. Bagger, N. Lambert, Phys.Rev.D79:025002 (2009)}
\lref\slava{A.M. Polyakov, V. Rychkov, Nucl.Phys.B594:272-286 (2001)}
\lref\brs{C. Becchi, A. Rouet, R. Stora, Annals Phys.98:287-321 (1976)}
\lref\progress{D. Polyakov, work in progress}

\centerline{\bf Introduction}

In our recent works ~{\ghost, \alfa, \selfgauge} we have shown that
RNS superstring theory is invariant under the set of global non-linear
space-time symmetries ($\alpha$-symmetries) that
mix matter and ghost degrees of freedom and enlarge the
space-time symmetry group pointing at their relation to hidden
space-time dimensions.
The generators of these symmetries ($\alpha$-generators)
can be classified in terms of superconformal ghost cohomologies
$H_n$ ~{\selfgauge}, with each ghost cohomology class contributing
its associate hidden dimension to the theory ~{\selfgauge}.
Typically, the $\alpha$-generators of $H_n$ ($n>0$) have the form:
\eqn\lowen{
L_n\sim\oint{{dz}\over{2i\pi}}e^{n\phi}F_{{{n^2}\over2}+n+1}(z)}
in the positive ghost number representation
(where  $\phi$  is the bosonized superconformal ghost field, and
$F_{{{n^2}\over2}+n+1}(z)$ are primary matter fields
of dimension ${{n^2}\over2}+n+1$
composed of RNS bosons and fermions $X$'s and $\psi$'s)
or
\eqn\lowen{
L_n\sim\oint{{dz}\over{2i\pi}}e^{-(n+2)\phi}F_{{{n^2}\over2}+n+1}(z)}
in the negative ghost cohomology representation $H_{-n-2}(n>0)$ ~{\selfgauge}.
The bosonic and fermionic ghost fields $\beta,\gamma,b$ and $c$
are bosonized
as usual according to ~{\fms}
$$\gamma(z)=e^{\phi-\chi}(z);
\beta(z)=e^{\chi-\phi}\partial\chi(z)\equiv\partial\xi{e^{-\phi}}(z)$$
and
$$c(z)=e^\sigma(z);b=e^{-\sigma}(z)$$
 
The important property of the $\alpha$-symmetries is that
there exist no analogues of global
space-time transformations induced by the operators (1)
at ghost numbers below $n$ (the analogues at pictures above $n$ 
can be obtained with the help of the picture-changing).
At the same time, the $\alpha$-transformations induced
by the generators (2) at the negative ghost numbers
have no analogues at pictures above $-n-2$;
the analogues at pictures $-n-3$ and below can be obtained
with the help of the inverse picture-changing.

From now on, for the rest of the paper we shall concentrate
on operators of the type (1) with positive ghost numbers.
The $\alpha$-generators with positive ghost numbers 
in the form (1) are typically not BRST-invariant
(they don't commute with supercurrent terms of the BRST charge)
and are therefore incomplete. That is, while they do generate the 
space-time symmetries, these symmetries are incomplete -
for example, the operators (1) do not act
on the $b-c$-ghost sector of the theory, while the full 
version $\alpha$-symmetries
have to include the $b-c$ ghosts as well.
Such a situation is similar to the well-known elementary example
of the BRST non-invariant Lorenz rotation generator in the RNS formalism

\eqn\lowen{L^{mn}=\oint{{dz}\over{2i\pi}}\psi^m\psi^n(z)}

where $\psi$'s are the RNS fermions:
although this symmetry operator  does generate the Lorenz
rotations for the fermions and fully  satisfies
the commutation relations of the rotation group, it does not
act on the RNS bosons (which of course must also be symmetric under the
space-time rotations), and such an incompleteness 
is directly related to the non-invariance of this operator.
Therefore, in order to restore the BRST-invariance of the 
$\alpha$-generators (1) (as well as that of the elementary generator
(3) of space-time rotations) one needs to add the correction
terms that also ensure (perhaps up to picture-changing) that
the improved BRST-invariant symmetry generators are complete,
 inducing the underlying global symmetries 
for all the relevant fields in the theory.
The correction terms can be obtained by the so called $K$-operator
procedure which is defined as follows.
Let $L=\oint{{dz}\over{2i\pi}}V(z)$ be some global symmetry
generator, incomplete (in the sense described above) and not BRST-invariant,
satisfying
\eqn\grav{\eqalign{\lbrack{Q_{brst}},V(z)\rbrack=\partial{U}(z)+W(z)}}
and therefore
\eqn\lowen{\lbrack{Q_{brst}},L{\rbrack}=\oint{{dz}\over{2i\pi}}W(z)}
where $V$ and $W$ are some operators of conformal dimension $1$
and $U$ is some operator of dimension zero.
Introduce the dimension 0 $K$-operator:
\eqn\lowen{K(z)=-4c{e}^{2\chi-2\phi}(z)\equiv{\xi}\Gamma^{-1}(z)}
satisfying
\eqn\lowen{\lbrace{Q_{brst}},K\rbrace=1}
where 
$\xi=e^\chi$ and $\Gamma^{-1}=4c\partial\xi{e^{-2\phi}}$ is the inverse 
picture-changing operator.
Suppose that the $K$-operator (6) has a non-singular OPE with $W(z)$:
\eqn\lowen{K(z_1)W(z_2)\sim{(z_1-z_2)^N}Y(z_2)+O((z_1-z_2)^{N+1})}
where $N\geq{0}$ and $Y$ is some operator of dimension $N+1$.
Then the complete BRST-invariant symmetry generator ${\tilde{L}}$
can be obtained from the incomplete non-invariant symmetry generator
$L$ by the following transformation:
\eqn\grav{\eqalign{
L\rightarrow{\tilde{L}}(w)=L+{{(-1)^N}\over{N!}}
\oint{{dz}\over{2i\pi}}(z-w)^N:K\partial^N{W}:(z)
\cr
+{1\over{{N!}}}\oint{{dz}\over{2i\pi}}\partial_z^{N+1}{\lbrack}
(z-w)^N{K}(z)\rbrack{K}\lbrace{Q_{brst}},U\rbrace}}
where $w$ is some arbitrary point on the worldsheet.
It is straightforward to check the invariance
of ${\tilde{L}}$ by using some partial integration along with
the relation (7) as well as the obvious identity
\eqn\lowen{\lbrace{Q_{brst}},W(z)\rbrace=
-\partial(\lbrace{Q_{brst}},U(z)\rbrace)}
that follows directly from (4).
The corrected invariant ${\tilde{L}}$-generators
are then typically of the form 
\eqn\lowen{{\tilde{L}}(w)=\oint{{dz}\over{2i\pi}}(z-w)^N{\tilde{V}}_{N+1}(z)}
(see the rest of the paper for the concrete examples)
with the conformal dimension $N+1$ operator ${\tilde{V}}_{N+1}(z)$
in the integrand satisfying
\eqn\lowen{{\lbrack}Q_{brst},{\tilde{V}}_{N+1}(z)\rbrack
=\partial^{N+1}{\tilde{U}}_0(z)}
where ${\tilde{U}}_0$ is some operator of dimension zero.
One unusual property of the invariant ${\tilde{L}}$-operators
is that, while being the generators of global symmetries in space-time,
they also appear to depend on an arbitrary point $w$ on a worldsheet,
except for the case $N=0$ (the latter case is particularly represented by the
space-time rotation generator as will be shown below).
It is easy to see, however, that
such a $w$-dependence is not in contradiction with 
the global properties of the ${\tilde{L}}$-generators as it
 will not appear in any correlation functions.
Indeed, since all the $w$ derivatives of ${\tilde{L}}(w)$ are
 BRST exact:
\eqn\grav{\eqalign{\partial_w^{k}{\tilde{L}}(w)
=\lbrace{Q_{brst}},\partial_w^{k-1}b_{-1}{\tilde{L}}(w)\rbrace
\cr
k=1,...,N}}
where $b$ is the b-ghost field and
$b_{-1}=\oint{{du}\over{2i\pi}}b(u)$,
changing the worldsheet location of ${\tilde{L}}$ leaves
it invariant up to BRST-trivial terms, that do not
contribute to correlators.
 For this reason, it is clear that the full OPE of any 
two ${\tilde{L}}(w)$'s is non-singular and all the OPE terms,
except for the contributions of the order of $(w_1-w_2)^0$,
are BRST exact:
\eqn\lowen{{\tilde{L}}_I(w_1){\tilde{L}}_J(w_2)
=C_{IJ}^K{\tilde{L}}_K(w_2)+\lbrack{Q_{brst}},...\rbrack}
Therefore the ${\tilde{L}}$-operators form the ground ring
with $C_{IJ}^K$ being the structure constants of the ring
(see the rest of the paper for concrete examples).
In this paper we shall particularly concentrate on
the properties of the $BRST$-exact derivatives of the ${\tilde{L}}$-operators.
These BRST-exact operators are of interest to us as they turn out to generate
new $local$ gauge symmetries on the worldsheet
(as opposed to global space-time 
symmetries induced by the ${\tilde{L}}$-operators 
themselves). In particular, consider the first derivative
of any  ${\tilde{L}}$-operator:
$\partial{\tilde{L}}=\lbrace{Q_{brst}},b_{-1}{\tilde{L}}\rbrace$.
This integral has conformal dimension 1 (accordingly the integrand
has overall
 conformal dimension 2) and turns out to generate local gauge symmetries
in RNS theory (demonstrated below in this paper).
Structurally, it is similar to the well-known conformal symmetries
generated by the integral of the dimension 2 stress-energy tensor
$\oint{T}(z)$ which is also BRST-exact, given by the commutator
of $Q_{brst}$ with the integral of the $b$-ghost.
As it is well-known, all the derivatives of the stress-energy tensor 
T in conformal field theory are the symmetry generators
as well, giving rise to infinite conformal symmetry in two dimensions.
Similarly, just as the first derivatives of ${\tilde{L}}$'s are the  
local symmetry generators, the higher order derivatives of
${\tilde{L}}(w)$ in $w$ induce local 
symmetries as well. The difference, however,
is that in this case one can have only the finite number
(up to $N$) of non-vanishing derivatives of ${\tilde{L}}$ (13).
So the gauge symmetries induced by the derivatives of the 
${\tilde{L}}$-operators are in fact finite-dimensional,
although, as will be demonstrated below, they do appear to have
Virasoro-like structure, as their commutators involve  solvable
subalgebras that look somewhat like a ``truncated'' Virasoro algebra.
Given the BRST-exact  generators of local symmetries $T_I$ 
(where $I$ labels the generators)
it is straightforward to construct (using the analogy with the stress-tensor)
the analogues $B_I$ 
of the $b$-ghost that is in the adjoint representation of the 
local gauge symmetry group. Then, if one is able to find  analogues
$C_I$ of the $c$-ghost which must be canonical conjugates of $B_I$,
it is straightforward to construct the nilpotent BRST operator
associated with these gauge symmetries. By definition,
the BRST operator is given by ~{\brs}
\eqn\lowen{Q={\sum_I}C^I{T}_I+{1\over2}\sum_{I,J,K}
f^{IJK}C_IC_JB_K}
where $f^{IJK}$ are the structure constants of the gauge symmetry
group. This operator is nilpotent, provided that
the $B_I$-ghosts are in the adjoint of the gauge symmetry group
and $f^{IJK}$ satisfy the Jacobi identities.
 In this paper, using the chain of the local gauge symmetries
inherited from the ground ring of the global $\alpha$-generators,
we shall construct a sequence of nilpotent 
BRST charges that can be classified in terms of the ghost cohomologies
(corresponding to the underlying $\alpha$-symmetries)
We conjecture that these new BRST operators (while technically
constructed of the currents in
the CFT of flat-dimensional superstring theory) describe RNS string
theories in various $curved$ backgrounds (particularly
defining the kinetic terms of string field theories built around the
curved backgrounds), although we leave the analysis of the cohomologies
of these operators for the future work which is currently in progress
~{\progress}.
The rest of the paper is organized as follows.
 In the section 2 we 
review the structure of $\alpha$-symmetry generators constructed in
 our previous works ~{\alfa, \selfgauge} and show them to form a ground ring.
The derivatives of the $\alpha$-generators are BRST-exact;
they induce local gauge symmetries mixing matter and ghost sectors
of RNS superstring theory and can be represented as BRST commutators
with certain non-minimal ghost  fields. 
In the section 3 we derive the generalized $B_I$ and $C_I$ ghosts
associated with these new gauge symmetries  for the superconformal
ghost number $1$ case
and construct the nilpotent BRST charge in the first
non-trivial ghost cohomology
$H_1$. In the section 4  we extend our construction
to higher ghost number generators, 
 deriving the associate $B-C$ ghost systems leading to nilpotent
BRST charges related to gauge
symmetries  derived from $\alpha$-generators at minimal ghost numbers
2 nad 3.
In the concluding section we discuss possible implications of our
results, particularly in relation to the problem of building
string field theories around curved backgrounds.
\centerline{\bf 1a. Ghost cohomologies: review of some definitions}
Before we start, we shall briefly review some definitions 
related to  ghost cohomologies (that are used to 
classify the $\alpha$-symmetries),
with some useful modifications (as compared to previous papers ~{\alfa,
 \selfgauge})
The $\alpha$-symmetry generators (1),(2) in the positive
and negative ghost picture representations have been often 
referred to as the elements of positive and negative ghost cohomologies
respectively. By definition, the negative
ghost cohomologies $H_{-n}(n\geq{3})$ consist of physical
(BRST-invariant and non-trivial) existing at minimal
negative picture $-n$ and the pictures below $(-n-1,-n-2,...)$, 
but not above $-n$.
The pictures $-n$ and below are related by the usual picture-changing
procedure; however, a picture-changing transformation applied to
such an operator at the minimal negative picture $-n$ annihilates
it, so there is no version of the elements of $H_{-n}$ at superconformal ghost
pictures above $-n$.  Example of elements of $H_{-n}$ 
are the $\alpha$-generators  (2) at pictire $-n$. 
Note that $n$ generally takes the values of $-3$
and below, as the cohomologies $H_{-1}$ and $H_{-2}$ are empty ~{\selfgauge}.
As for explicit examples, at this point we only have been able to identify
the elements of $H_{-n}$ up to $n=5$; for $H_{-n}$ at $n>5$ the 
expressions for the operators become quite cumbersome
and so far we have not been able to find them explicitly;
nevertheless at this time there is no evident reason to exclude the existence
of $H_{-n}$ for $n>5$. Next, 
the $positive$ ghost cohomologies $H_{n}$ (where $n>0$) consist
of physical (BRST-invariant and nontrivial) operators existing at
minimal $positive$ picture $n$ and above.
The pictures $n$ and above are related by the usual picture-changing
procedure; however, an inverse picture-changing transformation applied to
such an operator at the minimal positive picture $n$ annihilates
it, so there is no version of the elements of $H_{-n}$ at superconformal ghost
pictures below $n$. 
The isomorphic positive and negative
picture representations (1) and (2)
for the $\alpha$-generators hint that there is an similarity between the 
elements of $H_{n}$ and $H_{-n-2}$, conjectured in our previous works 
~{\alfa, \selfgauge}.
Things, however, appear to be a bit more subtle.
The negative picture representation (2) of the $\alpha$-generators 
is BRST-invariant and the symmetries generated by them are complete
 (up to picture changing). The positive picture expressions (1)
for the $\alpha$-symmetry generators are not, on the other hand,
BRST-invariant and are not complete (in the sense described above).
So they are not, strictly speaking, the elements of $H_n (n>0)$,
even though there exist no versions of the
incomplete $\alpha$-transformations below picture $n$.
In order to make them BRST-invariant
, one has to add the correction terms, using the $K$-operator procedure
(9). The $complete$ BRST-invariant expressions (9) for  the positive picture
$n$ $\alpha$-generators can be, on the contrary, transformed 
by the inverse picture changing operations to lower pictures 
(as will be demonstrated below),
although the ghost-matter mixing always persists and one is
never able to decouple the ghost and the matter degrees of freedom 
for such operators, even at superconformal picture zero.
So strictly speaking, it is not quite accurate to identify  the 
operators (1),(9) with the elements of positive ghost cohomologies.
Nevertheless for the sake of brevity and convenience 
 below in the text we shall sometimes refer to  operators,
 existing at minimal positive picture $n$ and above but not below $n$
(such as the ``abbreviated'' $\alpha$-generators (1)), as the ``elements
of $H_n$'' (in characters) even if such operators are not BRST-invariant.
At the same time, the complete positive 
picture $\alpha$-generators (9) can be described 
rigorously and more adequately in terms 
of the $b-c$ ghost cohomologies $R_{2n}$ that 
will be defined in the next section.

\centerline{\bf 2. Ground Ring of $\alpha$-Generators and  
 Associate Local Gauge Symmetries}

In string theory the global space-time symmetries are typically generated
by primary fields of conformal dimension 1 (commuting with
 BRST charge 
but not BRST-trivial), while local gauge symmetries are induced  by BRST
exact operators (that can have 
various conformal dimensions and are not necessarily primary) , 
given by commutators of BRST operator with appropriate ghost 
fields in the adjoint representation of the gauge group.
For example, the generators of local gauge symmetries on the worldsheet,
the stress-energy tensor $T$ and the supercurrent $S$, are the dimension
2 and $3\over2$ fields given by 
\eqn\lowen{T=\lbrace{Q_0},b\rbrace} and 
\eqn\lowen{S=\lbrack
{Q_0},\beta\rbrack,} 
while the dimension 1 primaries
\eqn\lowen{L^{m}=\oint{{dz}\over{2i\pi}}\partial{X^m}}
 and
\eqn\lowen{
L^{mn}=\oint{{dz}\over{2i\pi}}\psi^m\psi^n}
 generate Lorenz translations 
and rotations on the worldsheet.
Here
$$Q_0=\oint{{dz}\over{2i\pi}}(cT-bc\partial{c}-{1\over2}\gamma
\psi_m\partial{X^m}-{1\over4}b\gamma^2)$$
is the standard BRST operator in RNS theory, having
overall superconformal picture zero;
we have marked it with the 0 subscript, as below
we shall encounter the  sequence of alternative nilpotent
BRST operators $Q_n(n>0)$, existing at minimal positive superconformal
pictures other than zero (that appear to imitate RNS superstring
theories at various curved backgrounds).
Before we start discussing the $\alpha$-symmetries
and the associate local gauge symmetries, let us
consider (as a warm up example) the elementary case
of the space-time rotational symmetry in RNS theory.
As was noted above, the space-time rotation operator $L^{mn}$ 
is not  BRST-invariant and, as a consequence it is incomplete
as it generates rotational symmetry transformations only for $\psi$'s  
but not for $X$'s. 
 To construct a complete version of the generator
of Lorenz rotations, which acts both on $X$'s and $\psi$'s one needs
to improve $L^{mn}$ with $bc$-ghost dependent terms using the $K$-operator
prescription. In case of the rotation 
operator (19) the $K$-operator prescription (9)
gives $N=0$ and the complete BRST-invariant operator for the
space-time rotations is 
\eqn\grav{\eqalign{L^{mn}\rightarrow
{\tilde{L}}^{mn}=L^{mn}-2\oint{{dz}\over{2i\pi}}c\xi{e^{-\phi}}\partial
{X^{[m}}\psi^{n]}}}
It is then straightworward to check that, up to a picture-changing 
transformation, ${\tilde{L}}^{mn}$ generate the Lorenz rotations for both
$\psi$'s and $\partial{X}$'s, e.g.
\eqn\lowen{\Gamma\lbrack\epsilon^{mn}{\tilde{L}}_{mn}\partial{X^p}\rbrack
=\Gamma\epsilon_{pn}\partial{X^n}+\lbrack{Q_0,...}\rbrack}
Note that the BRST-invariant rotation operator ${\tilde{L}}^{mn}$
is outside the small Hilbert space, as it manifestly depends on $\xi$.
The next, far less trivial example of global space-time supersymmetries
in superstring theory is given by the hierarchy of $\alpha$-symmetries. 
These global space-time symmetries are realised non-linearly,
mixing the matter and the ghost sectors of RNS superstring theory and can be
classified in terms of ghost cohomologies. 
In our previous works ~{\alfa, \selfgauge}
 we have analyzed the properties of
these unusual symmetries, showing them to originate from hidden extra dimensions
of space-time, with each particular ghost cohomology subsector 
of $\alpha$-symmetries adding up an extra  dimension 
to space-time symmetry group. The $\alpha$-symmetry also turns out to play
an important role in the correspondence between strings and QCD dynamics,
as vertex operators for the octet of gluons  with field-theoretic
single pole structure of the scattering amplitudes, can be constructed
by $\alpha$-transforming a standard photon vertex operator
with SU(3) subgroup of $\alpha$-generators, which truncated
versions are characterized
by the lowest three
minimal superconformal ghost numbers $1,2$ and $3$
(given the minimal ghost number - extra dimension
correspondence, each of 3 hidden dimensions can be associated with particular
colour-anticolour pair in this approach ~{\selfgauge}).
The first and the simplest example of $\alpha$-symmetry is given by
the transformations which truncated generator is characterized 
by the minimal superconformal ghost number 1.
It can be checked that the full matter$+$ghost RNS action:
\eqn\grav{\eqalign{S_{RNS}=S_{matter}+S_{bc}+S_{\beta\gamma}\cr
S_{matter}={1\over{2\pi}}\int{d^2z}(\partial{X_m}\bar\partial{X^m}
+\psi_m\bar\partial\psi^m+{\bar\psi}_m\partial{\bar\psi}^m)\cr
S_{bc}={1\over{2\pi}}\int{d^2z}(b\bar\partial{c}+{\bar{b}}\partial
{\bar{c}})\cr
S_{\beta\gamma}={1\over{2\pi}}\int{d^2z}(\beta\bar\partial\gamma
+\bar\beta\partial{\bar\gamma})}}
is invariant under the following transformations (with $\alpha$ being
a global parameter):
\eqn\grav{\eqalign{\delta{X^m}=\alpha{\lbrace}2e^\phi\partial\psi^m+
\partial(e^\phi\psi^m)\rbrace\cr
\delta\psi^m=-\alpha{\lbrace}e^\phi\partial^2{X^m}+2\partial(e^\phi\partial
{X^m})\rbrace\cr
\delta\gamma={\alpha}
e^{2\phi-\chi}(\psi_m\partial^2{X^m}-2\partial\psi_m\partial
{X^m})\cr
\delta\beta=\delta{b}=\delta{c}=0}} so that
\eqn\grav{\eqalign{\delta{S_{matter}}=-\delta{S_{\beta\gamma}}
={1\over{2\pi}}\int{d^2z}(\bar\partial{e^\phi})(\psi_m\partial^2{X^m}
-2\partial\psi_m\partial{X^m})\cr
\delta{S_{bc}}=\delta{S_{RNS}}=0}}
The generator of the transformations (23) is given by
\eqn\grav{\eqalign{L_{\alpha+}=\oint{{dz}\over{2i\pi}}
e^\phi(\psi_m\partial^2{X^m}-2\partial\psi_m\partial{X^m})
\equiv\oint{{dz}\over{2i\pi}}e^{\phi}F(X,\psi)}}
where it is convenient 
to introduce the notation for the dimension ${5\over2}$ primary field:
\eqn\lowen{F(X,\psi)=
\psi_m\partial^2{X^m}-2\partial\psi_m\partial{X^m}} 
As in the case of the rotation generator, the integrand  of the $\alpha$
-generator (25) is a primary field of dimension 1, however it is not
BRST-invariant since it doesn't commute with the supercurrent terms of the
BRST charge; so similarly one  has to introduce the $bc$-dependent
correction terms to make it BRST-invariant, using the 
$K$-operator prescription (9).

The BRST-invariant extension of $L_{\alpha+}$ is constructed by using
the $K$-operator prescription (9) requiring 
(as it is easy to check) $N=2$. The  commutator of BRST charge 
$Q_0$ with the integrand of $L_{\alpha+}$ is straightforward to compute.
In addition to the dimension ${5\over2}$ primary field $F(X,\psi)$ introduced
above, it is also convenient to introduce the dimension 2 primary field
\eqn\lowen{L(X,\psi)=2\partial\psi_m\psi^m-\partial{X_m}\partial{X^m}}
Along with the matter stress tensor 
$T_{matter}=-{1\over2}\partial{X_m}
\partial{X^m}-{1\over2}\partial\psi_m\psi^m$
and the matter supercurrent $G=-{1\over2}\psi_m\partial{X^m}$ the $L$ and
$F$  matter primaries satisfy the following useful OPEœôòùs:

\eqn\grav{\eqalign{G(z)L(w)= {{F(w)}\over{{z-w}}}+...\cr
G(z)F(w)={{L(z)}\over{(z-w)^2}}+{1\over4}{{\partial{L(w)}}\over{z-w}}+...\cr
F(z)L(w)=-{{24G(w)}\over{(z-w)^3}}+{{4\partial{G(w)}-2F(w)}\over{(z-w)^2}}
+{{4\partial^2G(w)+4\partial{F(w)}-6\partial^2\psi_m\partial{X^m}}\over{z-w}}
+...\cr
F(z)F(w)=-{{22d}\over{(z-w)^5}}+ {{2L(w)+20T_{matter}(w)}\over{(z-w)^3}}
+{{\partial{L}(w)+10\partial{T_{matter}}(w)}\over{(z-w)^2}}\cr
+{1\over{z-w}}(3\partial^3{\psi_m}\psi^m(w)-6\partial^2\psi_m\partial{\psi^m}(w)
+6\partial^3{X_m}\partial{X^m}+
\partial^2{X_m}\partial^2{X^m})(w)+...
\cr
L(z)L(w)={{6d}\over{{(z-w)^4}}}+{1\over{(z-w)^2}}(16T_{matter}(w)-4L(w))
\cr+...
+{1\over{(z-w)}}(8\partial{T_{matter}}(w)-2\partial{L}(w))
}}
(where $d$ is the number of space-time dimensions)
Using the OPE's (28), 
the commutator of  $Q_0$ with the integrand of $L_{\alpha+}$
gives
\eqn\grav{\eqalign{\lbrace{Q_0},e^\phi{F(X(z),\psi(z))}\rbrace
=\partial(ce^\phi{F})+e^{2\phi-\chi}(FG-{1\over2}LP^{(2)}_{\phi-\chi}
-{1\over4}\partial{L}P^{(1)}_{\phi-\chi})
\cr+be^{3\phi-2\chi}
FP^{(1)}_{2\phi-2\chi-\sigma}}}
where the conformal weight $n$ polynomials $P^{(n)}_f(\phi_1(z),...,\phi_n(z))$
are  the generalized Hermite polynomials defined as

\eqn\lowen{ P^{(n)}_f(\phi_1(z),...,\phi_n(z))=e^{-f(\phi_1(z),...,\phi_n(z))}
{{\partial^n}\over{\partial{z}}}e^{f(\phi_1(z),...,\phi_n(z))}}
with $f$ being some function of the fields $\phi_1(z),œôó¦,\phi_n(z)$
(note that exponents in the definition are multiplied algebraically rather
than in terms of OPE). For example, for $P^{(1)}_{\phi-\chi}$ one has
$f(\phi,\chi)=\phi-\chi$ and therefore $P^{(1)}_{\phi-\chi}=
\partial\phi-\partial\chi$.
So we have to consider the $K$-operator prescription (9) with
 $U=ce^\phi{F}$, $W= e^{2\phi-\chi}(FG-{1\over2}LP^{(2)}_{\phi-\chi}
-{1\over4}\partial{L}P^{(1)}_{\phi-\chi})+be^{3\phi-2\chi}
FP^{(1)}_{2\phi-2\chi-\sigma}$ and, accordingly, $N=2$.
The straightforward evaluation of the OPE between $K$ 
and $\partial^2{W}$ gives:
\eqn\grav{\eqalign{:K\partial^2(e^{2\phi-\chi}(FG-{1\over2}LP^{(2)}_{\phi-\chi}
-{1\over4}\partial{L}P^{(1)}_{\phi-\chi})):
=-8c\xi((FG-{1\over2}LP^{(2)}_{\phi-\chi}
-{1\over4}\partial{L}P^{(1)}_{\phi-\chi}))\cr
-{1\over4}:L\partial^2(be^{3\phi-2\chi}FP^{(1)}_{2\phi-2\chi-\sigma}):
=\partial^2(e^\phi{F})-e^\phi
{F}P^{(2)}_{2\phi-2\chi-\sigma}\cr
-8c\xi(FG-{1\over2}LP^{(2)}_{\phi-\chi}
-{1\over4}\partial{L}P^{(1)}_{\phi-\chi})}}
so the full BRST-invariant extension of the $\alpha$-generator
$L_\alpha$ is given by
\eqn\grav{\eqalign{{\tilde{L}}_\alpha(w)
=\oint{{dz}\over{2i\pi}}(z-w)^2{\lbrack}{1\over2}e^\phi
{F}P^{(2)}_{2\phi-2\chi-\sigma}(z)+
\cr
4c\xi(FG-{1\over2}LP^{(2)}_{\phi-\chi}
-{1\over4}\partial{L}P^{(1)}_{\phi-\chi})-24\partial{c}{c}e^{2\chi-\phi}F
\rbrack}}
Note that, unlike the case of the rotation generator ($N=0$ case),
the full BRST-invariant expression ${\tilde{L}}_{\alpha+}(w)$
for the $\alpha$-generator,
obtained by the $K$-operator transformation (9) with $N=2$,
depends on an arbitrary worldsheet point $w$.
This is related to the fact that
${\tilde{L}}_{\alpha+}$ is an element of the $b-c$ 
ghost cohomology $R_2$, described below. Namely,
 the $b-c$ ghost cohomologies $R_N$
are defined as follows.
Recall that, while a $\beta-\gamma$ picture for
an RNS operator refers to its superconformal ghost number,
a $b-c$-picture $N$ ~{\self} for some operator $L$ refers 
to its  representation in the form $L=\oint{{dz}\over{2i\pi}}(z-w)^NV(z)$
with $V$ being a dimension $N+1$ operator satisfying
$\lbrack{Q_0},V\rbrack=\partial^{N+1}U$ where U is some operator of
dimension 0, so $L$ is invariant and $w$ is some arbitrary point
on a worldsheet. For example the $N=0$ case corresponds to 
standard integrated vertex operators in open string theory and $N=-1$ case
gives an unintegrated dimension 0 vertex operator located at $w$.
Just as $\beta-\gamma$ pictures can be changed by using the
picture-changing operator 

\eqn\grav{\eqalign{\Gamma(w)=:\delta(\beta)G(z):=:e^\phi{G}(w):
=-{1\over2}e^\phi\psi_m\partial{X^m}-{1\over4}be^{2\phi-\chi}
P^{(1)}_{2\phi-2\chi-\sigma}+c\partial\xi}}

where $G$ is the full matter$+$ghost worldsheet supercurrent,
the  operator changing the $b-c$ pictures 
can be obtained by the bosonic moduli
integration in the functional integral
for RNS amplitudes and is given by ~{\self}

 \eqn\lowen{Z(w)=:b\delta(T)(w)=
\oint{{dz}\over{2i\pi}}(z-w)^3\lbrack{bT}(z)+4c\partial\xi\xi
{e^{-2\phi}}T^2(z)\rbrack}.

Just as $\beta-\gamma$ ghost cohomology $H_N$ consists
of physical operators that exist at minimal $\beta-\gamma$ picture
N and above, but cannot be represented at pictures below $N$ by using
picture-changing transformation  with $\Gamma$ or 
its inverse, the $b-c$ ghost
cohomologies $R_N(N>0)$ 
consist of physical operators that exist at minimal $b-c$
picture $N$ or above, 
but cannot be related to pictures below $N$ through 
any transformation by $Z$. 
Thus the complete generator (20) of space-time rortations is the 
element of $R_0$, while the complete $\alpha$-generator
${\tilde{L}}_{\alpha+}$ is the element of
 $R_2$. In the following sections
we will also encounter the examples
of $\alpha$-generators of $R_4$ and $R_6$, along with their
associate gauge symmetries on the worldsheet leading to 
appearance of  new BRST charges.

\centerline{\bf 3. New Local Gauge Symmetries and New
BRST Charges}

It is straightforward to show that
the full BRST-invariant operator ${\tilde{L}}_{\alpha+}$ (32)
induces the full set of global space-time symmetries for the matter
and the ghost variables in RNS formalism (including the $b-c$ sector),
unlike the ``abbreviated'' operator $L_{\alpha+}$ (25) generating the truncated
version (23) of the $\alpha$-symmetry.
One seemingly unusual property of ${\tilde{L}}_{\alpha+}(w)$ is that, while
generating global space-time symmetry, it explicitly depends
on an arbitrary worldsheet coordinate $w$. This 
ambiguity, however, can be easily resolved if we note that
 all of the non-vanishing derivatives of ${\tilde{L}}_{\alpha+}$
in $w$ (that is, $\partial{\tilde{L}}_{\alpha+}(w)$
and $\partial^2{\tilde{L}}_{\alpha+}(w)$) are BRST-exact, therefore there is no
dependence on $w$  up to BRST-trivial  terms (e.g. an insertion of
${\tilde{L}}_{\alpha+}$ in any correlator won't depend on $w$)
It is not difficult to see that the relevant BRST commutators are given by:

\eqn\grav{\eqalign{
L^1_{\alpha+}\equiv\partial{\tilde{L}}_{\alpha+}(w) 
=\lbrace{Q_0},B_{\alpha+}(w)\rbrace\equiv
\lbrace{Q_0},b_{-1}{\tilde{L}}_\alpha(w)\rbrace\cr
L^2_\alpha\equiv\partial^2{\tilde{L}}_{\alpha+}(w)
=\lbrace{Q_0},{\partial}B_{\alpha+}(w)\rbrace}}
It is also helpful to have a manifest expression for $B_{\alpha+}(w)$.
Simple calculation gives

\eqn\grav{\eqalign{B_{\alpha+}(w)=\oint{{dz}\over{2i\pi}}
(z-w)^2\lbrace
-2b{e^\phi}{F}P^{(1)}_{2\phi-2\chi-\sigma}(z)
\cr
+8\xi(FG-{1\over2}LP^{(2)}_{\phi-\chi}
-{1\over4}\partial{L}P^{(1)}_{\phi-\chi})
+24\partial{c}e^{2\chi-\phi}F{\rbrace}}}

Next, it is straightforward to check that the dimension $1$ BRST-exact
operator $L^1_{\alpha+}$ (which integrand has conformal dimension 2)
generates {\it local} transformations for RNS fields that mix
the ghost and the matter sectors leaving $S_{RNS}$ invariant:
\eqn\grav{\eqalign{
\delta({\partial{X^m}}(w))=\epsilon(w){\lbrace}
-16\partial(c\xi{G})\psi^m-8c\xi{F}\psi^m
\cr
+8c\xi{P^{(2)}_{\phi-\chi}}\partial{X^m}
-4\partial(c\xi{P^{(1)}}_{\phi-\chi}\partial{X^m})
+2c\xi{P^{(1)}}_{\phi-\chi}\partial^2{X^m}
\cr
+2\partial((e^\phi{P^{(2)}_{2\phi-2\chi-\sigma}}-24\partial{c}c
\partial\xi\xi{e}^{-\phi})\psi^m)\rbrace\cr
+\partial\epsilon(w)\lbrace
-16c\xi{G}\psi^m-4c\xi{P^{(1)}}_{\phi-\chi}\partial{X^m}
\cr
+2(e^\phi{P^{(2)}_{2\phi-2\chi-\sigma}}-24\partial{c}c
\partial\xi\xi{e}^{-\phi})\psi^m\rbrace
\cr
\delta\psi^m=\epsilon(w){\lbrace}
-2(e^\phi{P^{(2)}_{2\phi-2\chi-\sigma}}-24\partial{c}c
\partial\xi\xi{e}^{-\phi})\partial{X^m}
+
16c\xi{G}\partial{X^m}
\cr
+8c\xi{P^{(2)}}_{\phi-\chi}\psi^m
-4\partial(c\xi{P^{(1)}_{\phi-\chi}}\psi^m)\rbrace
-4\partial\epsilon(w)c\xi{P^{(1)}_{\phi-\chi}}\psi^m
\cr
\delta{c}(w)=\epsilon(w)c{e^\phi}F(w)\cr
\delta{b}(w)=-24\epsilon(w)c\partial\xi\xi{e^{-\phi}}F(w)
\cr
\delta\beta(w)=\epsilon(w)(4\xi{F}+8c\partial\xi\xi{e^{-\phi}}L(w))
\cr
\delta\gamma=0}}
It is important to note that the symmetry transformations
(37) are induced by the generator $L^{1}_{\alpha+}(w)$
 applied to RNS variables
located at the $same$ point $w$. If RNS variables located
at $w$ are transformed by $L^1_{\alpha+}$ located at any point
other than $w$, $S_{RNS}$ is not symmetric under such transformations.

Next, it is a bit tedious but straightforward to check that
the ring of the gauge symmetry 
generators $L^1_{\alpha+}$ and $L^2_{\alpha+}$
is commutative, i.e. 
\eqn\lowen{\lbrack{L^1_{\alpha+}(w), L^2_{\alpha+}(w)}\rbrack=0}
and in addition 
$L^{(1,2)}_{\alpha+}(w)$ commute with the ghost variable $B_{\alpha+}(w)$,
as well as with all its non-vanishing derivatives
(that is, the first and the second derivatives in $w$).
Note that, while the dimension 1  
generator $L^1_{\alpha+}$ inducing local gauge transformations
at ghost number $1$-level
can be regarded as a counterpart of 
conformal transformations induced by
$T_{-1}=\oint{T(z)}$ at ghost number zero,
the commutation relations (38) indicate that the dimension 1
ghost variable $B_{\alpha+}$, associated with the gauge symmetries (37)
is the prototype of the integral of the usual $b$-ghost
$\oint{b}$, associated with the conformal transformations.
It is important to stress that
 that neither $L^1_{\alpha+}$ nor $B_{\alpha+}$ are related
to $\oint{T}$ and $\oint{b}$ by any picture-changing
(just as the global $\alpha$-symmetry transformations (23)
have no picture zero analogue ).
 The gauge symmetry  (37) is essentially different
from conformal symmetry and $L^1_{\alpha+}$ is a generator
physically different from $\oint{T}$.
Having the gauge symmetry  (37), its generator
$L^1_{\alpha+}$ and the associate ghost $B_{\alpha+}$ (analogue
of $\oint{b}$, the commutative ring of $L^{1,2}_\alpha$
(which elements also commute with $B_{\alpha+}$ and its derivatives) 
the only missing component to construct a nilpotent 
BRST charge at the $H_1$-level, related 
to the gauge symmetry (37), is the analogue
$C_{\alpha+}$ of the $c$-ghost, conjugate to $B_{\alpha+}$.
Since $b$ and $c$ ghosts satisfy the canonical relation

\eqn\lowen{\lbrace{\oint{b},c}\rbrace=1,}

the $B_{\alpha+}$ and $C_{\alpha+}$ ghosts have to satisfy
\eqn\lowen{\lbrace{B_{\alpha+},C_{\alpha+}}\rbrace=1}
The object that is able to satisfy the anticommutator
(40) with $B_{\alpha+}$ must be a $local$ field of conformal
dimension $-1$. Since the usual $c$-ghost is also a primary field,
our goal now is  to construct the conformal dimension $-1$
primary field satisfying (40). Since the expression (36) for
$B_{\alpha+}$ is at picture $1$, $C_{\alpha+}$ satisfying (40) must be at picture
$-1$. However, simple structural and dimensional analysis
shows that there exist no picture $-1$-object with such properties.
For this reason, we shall replace the canonical relation (40)
with
\eqn\lowen{{\lbrace}{\Gamma^{-2}}B_{\alpha+},C_{\alpha+}\rbrace=\Gamma}
with $\Gamma=\lbrace{Q_0},\xi\rbrace$ being the direct
picture-changing operator or, equivalently, the picture-transformed 
unit operator and $\Gamma^{-2}=:\Gamma^{-1}\Gamma^{-1}:$ being the square
of the inverse picture-changing $\Gamma^{-1}=4c\partial\xi{e^{-2\phi}}$
satisfying $:\Gamma\Gamma^{-1}:=1$.

 Note that, even though $B_{\alpha+}$ is off-shell,
picture-changing transformations (both the direct and the inverse)
are still well-defined for it: since
the $L_{\alpha+}^1$ is on-shell by construction, 
 it can be picture-transformed in a well-defined manner; then,
since $\Gamma^{-2}$ is invariant, by 
definition 
\eqn\lowen{\Gamma^{-2}L_{\alpha+}^1=
\lbrace{Q_0},B_{\alpha+}^{(-2)}\rbrace
\equiv\lbrace{Q_0},{\Gamma^{-2}}B_{\alpha+}\rbrace}
Applying the inverse picture-changing twice with $\Gamma^{-1}$
it is straightforward to obtain the picture-transformed expression
for  $B_{\alpha+}$ at ghost number $-1$:
\eqn\grav{\eqalign{
B_{\alpha+}^{(-1)}(w)\equiv\Gamma^{-2}B_{\alpha+}(w)
=\oint{{dz}\over{2i\pi}}(z-w)^2\lbrace-8
\partial{c}{c}
e^{3\phi-4\chi}\cr\times
\lbrace{{1\over2}P^{(2)}_{-\sigma}}
\lbrack
-{{3}\over8}\partial^2L+{1\over4}\partial{L}
P^{(1)}_{-16\phi+3\chi-3\sigma}
\cr+L 
\times(-{3\over2}\partial^2\phi+{{11}\over8}\partial^2\chi
+{3\over8}\partial^2
\sigma-4(\partial\phi)^2
+{5\over8}(\partial\chi)^2
\cr
+{1\over8}
(\partial\sigma)^2+6\partial\phi\partial\chi
-{1\over2}\partial\phi\partial
\sigma+{7\over4}\partial\chi\partial\sigma)\rbrack
\cr
+{1\over6}P^{(3)}_{-\sigma}(-{3\over4}\partial{L}+L({1\over4}\partial\sigma
-{1\over2}\partial\phi))+{1\over{48}}P^{(4)}_{-\sigma}L\rbrace
\cr
-ce^{2\chi-3\phi}\lbrace{P^{(1)}_{-\sigma}}  
\times\lbrack
{-{3\over8}\partial^2{F}-{1\over4}\partial{F}{P^{(1)}_{\phi-2\chi+2\sigma}}}
\cr
+F{\lbrack}{1\over8}\partial^2\phi+{{15}\over4}\partial^2\chi
-{1\over4}\partial^2\sigma
+{{13}\over8}(\partial\phi)^2=
\cr
-3(\partial\chi)^2-{5\over2}\partial\phi
\partial\chi
-{3\over2}\partial\phi\partial\sigma+\partial\chi\partial\sigma
\rbrack 
\cr
+{1\over2}{P^{(2)}_{-\sigma}}({-1\over2}\partial{F}
+F(-{3\over2}\partial\phi-\partial\chi))
-{1\over{24}}P^{(3)}_{-\sigma}\rbrace(z)}}
Next, it is straightforward to check that the conformal dimension $-1$
primary field $C_{\alpha+}$, satisfying the picture-transformed canonical
relation (40) with $B_{\alpha+}$ is given by:

\eqn\grav{\eqalign{
C_{\alpha+}(w)
={1\over2}e^{3\phi-\chi}\lbrace
F({1\over3}P^{(3)}_{\phi-\chi}+{1\over2}\partial\phi{P^{(2)}_{\phi-\chi}})
+GL({1\over2}P^{(2)}_{\phi-\chi}+\partial\phi{P^{(1)}_{\phi-\chi}}
+{1\over2}\partial{F}P^{(2)}_{\phi-\chi})
\cr
+\partial{G}LP^{(1)}_{2\phi-\chi}+G\partial{L}P^{(1)}_{\phi-\chi}
+{1\over2}\partial^2{G}L+\partial{G}\partial{L}\rbrace
\cr
+be^{4\phi-2\chi}\lbrace
{1\over2}GFP^{(1)}_{\phi-\chi-{3\over4}\sigma}P^{(1)}_{\phi-\chi}
+{1\over{12}}LP^{(3)}_{\phi-\chi}P^{(1)}_{\phi-\chi-{3\over4}\sigma}
+{1\over{16}}\partial{L}P^{(2)}_{\phi-\chi}P^{(1)}_{\phi-\chi-{3\over4}\sigma}
\rbrace
\cr
+\partial{b}{b}e^{5\phi-3\chi}\lbrace
-{1\over8}P^{(1)}_{\phi-\chi-{3\over4}\sigma}P^{(2)}_{2\phi-2\chi-\sigma}
+{1\over{32}}P^{(3)}_{2\phi-2\chi-\sigma}\rbrace}}
The expression (44) for $C_{\alpha+}$ is at the ``picture cohomology $H_2$''
i.e. it has no analogues at pictures below 2 - in particular,
that's the reason why
one has  to transform $B_{\alpha+}$ in order to satisfy the 
canonical relation (41).
Having the gauge symmetry generators $L_{\alpha+}^{1,2}$, the associate
$B_{\alpha+}$ and $C_{\alpha+}$-ghosts, as well as the commutation 
relations (38),
it is now straightforward to construct the 
nilpotent BRST-charge in the $H_1$ ghost
cohomology which, by definition, is given by
\eqn\grav{\eqalign{
Q_1={{C_{\alpha+}}}^1L_{\alpha}^1+{{C_{\alpha+}}}^2L_\alpha^2}}
where
\eqn\grav{\eqalign{
{{C_{\alpha+}}}^1\equiv{C_\alpha+},
\cr
{{C_{\alpha+}}}^2
=\oint{{dz}\over{2i\pi}}C_{\alpha+}(z)}}

To construct the manifest expression for $Q_1$ in terms of an
 integral of BRST current, it is convenient
 transform the full BRST-invariant expression for
${{{\tilde{L}}_{\alpha+}}}(w)$ (as well as those
for $L^{1}_{\alpha+}=\partial{{{\tilde{L}}_{\alpha+}}}(w)$
and $L^2_{\alpha+}=\partial^2{{{\tilde{L}}_{\alpha+}}}(w)$)
to $-1$ picture, as $C_{\alpha+}$ only exists at picture $2$ and above,
and $Q_1$ should have minimal superconformal picture $1$.
Applying $\Gamma^{-2}$ to 
 ${{{\tilde{L}}_{\alpha+}}}(w)$ (32) twice,
it is straightforward to obtain

\eqn\grav{\eqalign{{{{\tilde{L}}_{\alpha+}}}^{(-1)}(w)
=
\oint{{dz}\over{2i\pi}}(z-w)^2\lbrace
-8\partial^2{c}\partial{c}c{e^{3\chi-4\phi}}
{\lbrack}-{3\over8}\partial^2L
\cr
+\partial{L}(-4\partial\phi
+3\partial\chi-{3\over4}\partial\sigma)
\cr
+L(-{3\over2}\partial^2\phi+{{11}\over8}\partial^2\chi+{3\over8}
\partial^2\sigma-4(\partial\phi)^2+{5\over8}(\partial\chi)^2
\cr
+{1\over8}
(\partial\sigma)^2+6\partial\phi\partial\chi
-{1\over2}\partial\phi\partial\sigma+{7\over4}\partial\chi\partial\sigma)\rbrack
\cr
-\partial{c}c{e^{2\chi-3\phi}}
\lbrack
-{3\over8}\partial^2F+\partial{F}(-{1\over4}\partial\phi+{1\over2}\partial\chi
-{1\over2}\partial\sigma)\cr
+F({1\over8}\partial^2\phi+{{15}\over4}\partial^2\chi-{1\over4}\partial^2\sigma
+{{13}\over8}(\partial\phi)^2
\cr
-3(\partial\chi)^2-{5\over2}\partial\phi
\partial\chi-{3\over2}\partial\phi\partial\sigma+\partial\chi\partial\sigma)
\rbrack\rbrace}}
Substituting (44) and (47) into the BRST charge (45) and 
evaluating the relevant
OPE's we obtain, upon a straightforward although somewhat lengthy 
computation, the remarkably simple expression:
\eqn\lowen{
Q_1=\oint{{dz}\over{2i\pi}}{\lbrace}ce^\phi{FP^{(1)}_{\phi-\chi}}
-{1\over8}e^{2\phi-\chi}(LP^{(2)}_{2\phi-2\chi-\sigma}+2GF)
-\partial{c}c\xi{L}\rbrace(z)}
This BRST charge is nilpotent by construction and its nilpotence can also be 
checked by direct computation.
It is an on-shell operator commuting with $Q_0$ and is an
element of $H_1$ ghost cohomology, i.e. it is unrelated
to the standard BRST charge $Q_0$ by any picture-changing transformation.
As was stressed above, the charge $Q_1$  (48) is an independent
BRST charge, originating from the ground ring of $\alpha$-symmetries,
with independent cohomology of physical states.

\centerline{\bf 4. $\alpha$-Generators of Higher Ghost Cohomologies: a Review}

For uncompactified critical RNS theory, $Q_1$ of
$H_1$ is the only additional
BRST charge present, as for this case there exist only one 
global $\alpha$-symmetry with two associate generators 
of local gauge symmetries deduced from ${\tilde{L_{\alpha+}}}$.
In non-critical or compactified cases, however, there are additional
$\alpha$-generators, at minimal positive ghost numbers $n>1$
~{\selfgauge}, due to interactions with the Liouville mode or
the compactified dimensions.
For completeness, below we shall give a short review of the basic properties
of the $\alpha$-generators at higher minimal
ghost numbers, briefly summarizing the results
of ~{\selfgauge}.

In general, for a {$d$-dimensional RNS theory}, 
there exist {$d+1$ additional $\alpha$-symmetries 
in $b-c$ ghost cohomology $R_2$} (with their
truncated non-invariant versions (1) having
 minimal superconformal
ghost number 1 prior to adding the correction terms).
For the brevity, below we shall give the 
 truncated  expressions 
for these operators (i.e. before adding the $b-c$ ghost correction terms);
the $K$-operator procedure is performed for these operators 
quite similarly to the $L_{\alpha+}\rightarrow{\tilde{L}}_{\alpha+}$ case
explained above. The remaining $d+1$ truncated $\alpha$-generators of $H_1$ 
(one $d$-vector and one Lorenz scalar) are
given by

\eqn\grav{\eqalign{L_{m\alpha}=\oint{{dz}\over{2i\pi}}
e^{\phi}\lbrace\partial^2\varphi\psi^m
-2\partial\varphi\partial\psi^m+\partial^2{X^m}\lambda
-2\partial{X^m}
\partial\lambda\rbrace}}
and
\eqn\lowen{L_{\alpha-}=\oint{{dz}\over{2i\pi}}e^{\phi}
\lbrace\partial^2\varphi\lambda-
2\partial\varphi\partial\lambda\rbrace}

where $\varphi$ and $\lambda$ are the components of the super Liouville 
field.

Combined with
{ ${{(d+1)(d+2)}\over{2}}$} Poincare symmetries
(including the Liouville direction), these ${d+2}$
ghost-matter mixing symmetries of $R_2$ enlarge space-time symmetry group
from $SO(2,d)$ to $SO(2,d+1)$, bringing in the first extra-dimension

Next, the $R_4$ cohomology can be shown to contain
$(d+3)$  $\alpha$-symmetries (with their truncated versions
having minimal superconformal ghost number 2)
which, combined with Poincare symmetries and $\alpha$-symmetries
of $H_2$ enlarge the space-time symmetry group
to $SO(2,d+2)$, bringing in  the second extra-dimension.
The truncated expressions for the $\alpha$-generators of
$R_4$ are given by
\eqn\grav{\eqalign{
L_{\beta{+}}=\oint{{dz}\over{2i\pi}}e^{2\phi}F_1(X,\psi)
F_1(\varphi,\lambda)(z)\cr
L_{\beta-}=-\oint{{dz}\over{2i\pi}}{e^{2\phi}}
F_{1m}(X,\lambda)F_1^m(\varphi,\psi)(z)\cr
L_{\beta{m}}=\oint{{dz}\over{2i\pi}}e^{2\phi}
(F_1^m(X,\lambda)F_1(\varphi,\lambda)
-F_1(X,\psi)F_1^m(\varphi,\psi))(z)\cr
L_{\alpha\beta}=\oint{{dz}\over{2i\pi}}
e^{2\phi}({1\over2}F_2(\lambda,\varphi)+
L_1(X,\psi)\partial{L_1}(\varphi,\lambda)-\partial{L_1}(X,\psi)
L_1(\varphi,\lambda))(z)}}
with  the matter$+$Liouville structures $L$ and $F$
($L_1,F_1$ and $F_1^m$) being the primary fields of dimensions 2 and $5\over2$:
\eqn\grav{\eqalign{F_1(X,\psi)=
\psi_m\partial^2{X^m}-2\partial\psi_m\partial{X^m}\cr
F_1(\varphi,\lambda)=\lambda\partial^2\varphi-
2\partial\lambda\partial\varphi\cr
F_1^m(X,\lambda)=\lambda\partial^2{X^m}-2\partial\lambda\partial{X^m}\cr
F_1^m(\varphi,\psi)=\psi^m\partial^2\varphi-2\partial\psi^m\partial\varphi\cr
L_1(X,\psi)=\partial{X_m}\partial{X^m}-2\partial{\psi_m}\psi^m\cr
L_1(\varphi,\lambda)=(\partial\varphi)^2-2\partial\lambda\lambda}}
and $F_2(\lambda,\varphi)$ being the primary field of dimension 5:
\eqn\grav{\eqalign{
F_2(\varphi,\lambda)
={1\over4}(\partial\varphi)^5
-{3\over4}\partial\varphi(\partial^2\varphi)^2+{1\over4}
(\partial\varphi)^2\partial^3\varphi\cr+
\lambda\partial\lambda(\partial^3\varphi-(\partial\varphi)^3)
-{3\over2}\lambda\partial^2\lambda\partial^2\varphi
+3\partial\lambda\partial^2\lambda\partial\varphi
\rbrace\cr\equiv
i:(\oint{e^{-i\varphi}\lambda})^3{e^{3i\varphi}\lambda}:}}
Finally, the  $d+4$ $\alpha$-generators of $R_6$
(with their truncated versions having
minimal superconformal ghost number 3) include one Lorenz
$d$-vector and 4 scalars,
 enlarging the symmetry
group to $SO(2,d+3)$, bringing in the third hidden dimension.
The truncated $\alpha$-generators of $R_6$ are given by
\eqn\grav{\eqalign{L_{\gamma{+}}=\oint{{dz}\over{2i\pi}}
e^{3\phi}{\lbrace}
2\partial{F_1}(X,\psi){F_2}(\varphi,\lambda)
-F_1(X,\psi)\partial{F_2}(\varphi,\lambda)\rbrace
\cr
L_{\gamma{m}}=\oint{{dz}\over{2i\pi}}e^{3\phi}\lbrace
2F_2^m(\psi,\lambda,\varphi)\partial{F_1}(X,\psi)
-\partial{F_2}(\psi,\lambda,\varphi)F_1(X,\psi)\cr
+2F_2(\varphi,\lambda)\partial{F_1^m}(X,\lambda)
-\partial{F_2}(\varphi,\lambda)F_1^m(X,\lambda)\rbrace\cr
L_{\gamma-}=\oint{{dz}\over{2i\pi}}{e^{3\phi}}
\lbrace
2G_2(\psi,\lambda,\varphi)\partial{F_1}(X,\psi)
-\partial{G_2}(\psi,\lambda,\varphi){F_1}(X,\psi)\cr
+3F_{2m}(\psi,\lambda,\varphi)\partial{F_1^m}(X,\lambda)
-2\partial{F_{2m}}(\psi,\lambda,\varphi)F_1^m(X,\lambda)
-\partial{F_2}(\lambda,\varphi)F_1(X,\psi)\rbrace
\cr
L_{\gamma\beta}=\oint{{dz}\over{2i\pi}}
e^{3\phi}{\lbrace}F_3(\varphi,\lambda)
+\partial{L_1}(X,\psi)L_2(\varphi,\lambda)-{4\over{11}}L_1(X,\psi)
\partial{L_2}(\varphi,\lambda)\rbrace\cr
L_{\gamma\alpha}=\oint{{dz}\over{2i\pi}}e^{3\phi}
{L_{2m}}(\varphi,\psi)L_1^m(X,\lambda)}}
with the additional matter$+$Liouville blocks given by:
\eqn\grav{\eqalign{F_2^m(\psi,\lambda,\varphi)=
\partial^2\psi^m\lambda\partial^2\varphi-\psi^m\partial^2\lambda
\partial^2\varphi
+3\partial^2\psi^m\partial\lambda\partial\varphi
-3\partial\psi^m\partial^2\lambda\partial\varphi\cr
G_2(\psi,\lambda,\varphi)=
4\partial\psi_m\partial^2\psi^m\partial\varphi-2\psi_m\partial^3\psi^m
\partial\varphi+(2d-4)(\lambda\partial^3\lambda\partial\varphi-
2\partial\lambda\partial^2\lambda\partial\varphi)\cr
L_2(\varphi,\lambda)=-{5\over4}(\partial\varphi)^4\partial\lambda
+{3\over4}(\partial^2\varphi)^2\partial\lambda
+{3\over2}\partial\varphi\partial^2\varphi\partial^2\lambda
-{5\over2}\partial\varphi\partial^3\varphi\partial\lambda\cr
-{1\over4}(\partial\varphi)^2\partial^3\lambda
-4\partial\varphi\partial^2\varphi\partial^2\lambda
+\partial^2\varphi\partial^3\varphi\lambda\cr
L_2^m(\varphi,\psi)=-{5\over4}(\partial\varphi)^4\partial\psi^m
+{3\over4}(\partial^2\varphi)^2\partial\psi^m
+{3\over2}\partial\varphi\partial^2\varphi\partial^2\psi^m
-{5\over2}\partial\varphi\partial^3\varphi\partial\psi^m\cr
-{1\over4}(\partial\varphi)^2\partial^3\psi^m
-4\partial\varphi\partial^2\varphi\partial^2\psi^m
+\partial^2\varphi\partial^3\varphi\psi^m\cr
L_1^m(X,\lambda)=\partial^2\lambda\psi^m+\lambda\partial^2\psi^m\cr
F_3(\varphi,\lambda)=:(\oint{e^{-i\varphi}\lambda})^4
{e^{4i\varphi}\lambda}:}}
Although the $\alpha$-symmetry
generators of $R_{2n}$ $b-c$ ghost cohomologies
(along with their associate rings of local gauge symmetries)
have not yet been constructed explicitly for $n>3$ cases
(as the manifest expressions for the $\alpha$-symmetry generators
become extremely cumbersome at higher $n$'s), it seems 
plausible that the $\alpha$-symmetries also exist at $n>3$ levels
as well, with each subset of the generators from $R_{2n}$ at a given $n$
adding the associate hidden space-time dimension 
(checked explicitly for $n=1,2,3$ ~{\selfgauge}).
There are both Lorenz scalars and $d$-vectors among
 $\alpha$-symmetry generators of the first three
cohomologies $n\leq{3}$; the scalar generators
have been shown to form the  $SU(3)$ subgroup.
If one takes an open string photon and acts on it with
the $\alpha$-generators of $SU(3)$ subgroup, one obtains
an $SU(3)$ octet of vertex operators of gluons
which tree level amplitude reproduces that of 
 $SU(3)$ QCD, i.e. this amplitude has a field-theoretic
rather than a stringy structure (with only the massless poles present).
The absence of an infinite tower
of massive poles (typical for 
 standard Veneziano amplitudes) is related to the
remarkable property of the $\alpha$-transformations proven
in ~{\selfgauge}: if applied to massless states,
they produce new physical massless vertex operators;
however, the $\alpha$-transformations applied  to any massive intermediate
 states
produce BRST-trivial operators that of course 
do not contribute  to the scattering amplitude. Thus the $\alpha$-symmetry
``erases'' the massive poles, to ensure the
proper field-theoretic behaviour of the gluon amplitude. 
 Higher order $\alpha$-generators may also be constructed in critical
RNS theory, e.g. by compactifying one of ten dimensions on $S^1$
and replacing the super Liouville mode with the compactified
dimension (and its worldsheet superpartner) in the expressions for the 
generators of $R_2,R_4$ and $R_6$. 

\centerline{\bf  5. $\alpha$-Generators of Higher Ghost Cohomologies:
Gauge Symmetries and BRST Charges}

Given the truncated versions of the $\alpha$-generators
of $R_2$, $R_4$ and $R_6$, their complete BRST-invariant 
expressions are straightforward to construct by using the $K$-operator
procedure (9). The construction is totally similar to the case
of ${\tilde{L}}_{\alpha+}$ of $R_2$, described above.
Upon the $K$-operator transformation, all
the  truncated $\alpha$-generators (49), (50)
of ghost number 1 become the elements
of the $b-c$ ghost cohomology $R_2$ (similarly
to the case of $L_{\alpha+}\rightarrow{\tilde{L}}_{\alpha+}$) 
considered above).
Next, it is straightforward to find out 
that, upon the $K$-operator transformation
(9) all the  truncated $\alpha$-generators (51)
of ghost number 2 become the elements
the  cohomology $R_4$, while the truncated operators (54) of 
ghost number 3 become the elements of
$R_6$.
 That is, 
in the case of
 the truncated superconformal ghost number $2$ $\alpha$-symmetries (51)
 the $K$-operator prescription  (9)
requires $N=4$ and is given by

\eqn\grav{\eqalign{L^{(2)}\rightarrow
{\tilde{L}}^{(2)}_{R_4}(w)=L^{(2)}-{1\over{24}}
\oint{{dz}\over{2i\pi}}(z-w)^4{L}\partial^4{W^{(2)}}(z)
\cr
+{1\over{24}}\oint{{dz}\over{2i\pi}}\partial_z^5{\lbrack}
(z-w)^4L\rbrack{L}{\lbrace}Q_0,U^{(2)}\rbrace}}
where the operators $W^{(2)}$ and $U^{(2)}$ are defined by the BRST
($Q_0$) commutator with the integrands of the truncated 
expressions (51) for the subset of the $\alpha$-
 generators $L^{(2)}$ of $H_2$:
\eqn\grav{\eqalign{L^{(2)}\equiv\oint{{dz}\over{2i\pi}}V^{(2)}
\cr
\lbrack{Q_0},V^{(2)}\rbrack=\partial{U^{(2)}}+W^{(2)}}}
Similarly,
for the superconformal ghost number $3$ truncated $\alpha$-symmetries (54)
 the $K$-operator prescription requires
$N=6$ and  is given by

\eqn\grav{\eqalign{L^{(3)}\rightarrow
{\tilde{L}}^{(3)}_{R_4}(w)=L^{(2)}-{1\over{{720}}}
\oint{{dz}\over{2i\pi}}(z-w)^6L\partial^6{W^{(3)}}(z)
\cr
+{1\over{{720}}}\oint{{dz}\over{2i\pi}}
\partial_z^7{\lbrack}
(z-w)^6L\rbrack{L}{\lbrace}Q_0,U^{(3)}\rbrace}}
where, as previously,
\eqn\grav{\eqalign{L^{(3)}\equiv\oint{{dz}\over{2i\pi}}V^{(3)}
\cr
\lbrack{Q_0},V^{(3)}\rbrack=\partial{U^{(3)}}+W^{(3)}}}
Structurally, the complete BRST ($Q_0$)-invariant
$\alpha$-generators ${\tilde{L}}^{(k)} (k=1,2,3)$ 
in the $b-c$ ghost
cohomologies $R_2,R_4$ and $R_6$ respectively, have the form:
\eqn\grav{\eqalign{
{\tilde{L}}^{(k)}(w)=\oint{{dz}\over{2i\pi}}(z-w)^{2k}{{\tilde{V}}^{(2k+1)}}
\cr
\lbrack{Q_0},{{\tilde{V}}^{(2k+1)}}\rbrack=\partial^{2k+1}{\tilde{U_k}}^{(0)}}}
where 
${{\tilde{V}}^{(2k+1)}}$ are the integrands of conformal dimension
$2k+1$ and ${\tilde{U_k}}^{(0)}$  are some operators of 
conformal dimension zero.
The ghost-matter structures of the operators  
 ${\tilde{L}}^{(k)} (k=1,2,3)$ in $R_{2k}$ respectively
are given by 
\eqn\grav{\eqalign{
{\tilde{L}}^{(1)}(w)=\oint{{dz}\over{2i\pi}}{(z-w)^2}
(e^\phi{R^{({9\over2})}}+ce^\chi{R^{(4)}}+\partial{c}ce^{2\chi-\phi}
R^{({5\over2})})(z)\cr
{\tilde{L}}^{(2)}(w)=\oint{{dz}\over{2i\pi}}(z-w)^4
(e^{2\phi}R^{(9)}+ce^{\phi+\chi}R^{({{15}\over2})}
+\partial{c}c{e^{2\chi}}R^{(5)})(z)\cr
{\tilde{L}}^{(3)}(w)=\oint{{dz}\over{2i\pi}}(z-w)^6
(e^{3\phi}R^{({{29}\over2})}+ce^{\chi+2\phi}R^{(12)}
+\partial{c}c{e^{\phi+2\chi}}R^{({{17}\over2})})(z)}}
where $R^{(k)}$ are the operators of conformal 
dimension $k$, made of matter fields and various polynomial
functions of ghost number currents, such as
$P^{(i)}_{\phi-\chi}$ and $P^{(j)}_{2\phi-2\chi-\sigma}$
with various $i$ and $j$.
For the sake of completeness, below we shall give the manifest
expressions for some most typical operators
${\tilde{L}}^{(k)}(w)$ for $k=2$ and $3$:
\eqn\grav{\eqalign{
{\tilde{L}}_{\beta+}(w)
=\oint{{dz}\over{2i\pi}}(z-w)^4\lbrace
{1\over{24}}{e^{2\phi}}F(X,\psi)F(\varphi,\lambda)
P^{(4)}_{2\phi-2\chi-\sigma}
\cr
+{1\over6}c{e}^{\chi+\phi}{\lbrack}
\partial{G}F(X,\psi)F(\varphi,\lambda)
+GF(X,\psi)F(\varphi,\lambda)P^{(1)}_{\phi-\chi}
+{1\over6}P^{(3)}_{\phi-\chi}(L(X,\psi)F(\varphi,\lambda)
\cr
-F(X,\psi)L(\varphi,\lambda))
+{1\over8}P^{(2)}_{\phi-\chi}({\partial}L(X,\psi)F(\varphi,\lambda)
-F(X,\psi){\partial}L(\varphi,\lambda))\rbrack
\cr
+20\partial{c}c\partial\xi\xi{F(X,\psi)}F(\varphi,\lambda)\rbrace}}
for $k=2$ and
\eqn\grav{\eqalign{{\tilde{L}}_{\gamma+}={1\over{720}}\oint{{dz}\over{2i\pi}}
(z-w)^6\lbrace
{e^{3\phi}}(2\partial{F_1}(X,\psi)F_2(\varphi,\lambda)
-F_1(X,\psi)\partial{F_2}(\varphi,\lambda))P^{(6)}_{2\phi-2\chi-\sigma}\rbrace
\cr
+\oint{{dz}\over{2i\pi}}
(z-w)^6\lbrace
c\xi{e^{2\phi}}{\lbrack}
(2GP^{(2)}_{\phi-\chi}+4\partial{G}P^{(1)}_{\phi-\chi}
+2\partial^2{G})
\cr\times
(2\partial{F_1}(X,\psi)F_2(\varphi,\lambda)
-F_1(X,\psi)\partial{F_2}(\varphi,\lambda))
\cr
+P^{(3)}_{\phi-\chi}
({1\over3}\partial^2{L_1}(X,\psi)F_2(\varphi,\lambda)
+{2\over3}{\partial}F_1(X,\psi)\partial{L_2}(\varphi,\lambda)
\cr
-{1\over6}\partial{L_1}(X,\psi){\partial}F_2(\varphi,\lambda)
-{4\over3}\partial{F_1}(X,\psi){L_2}(\varphi,\lambda))
\cr
+P^{(4)}_{\phi-\chi}({5\over{12}}
\partial{L_1}(X,\psi)F_2(\varphi,\lambda)
+{1\over6}F_1(X,\psi){L_2}(\varphi,\lambda)
\cr
-{1\over6}{L_1}(X,\psi){\partial}F_2(\varphi,\lambda))
+{2\over{15}}P^{(5)}_{\phi-\chi}
2L_1(X,\psi)F_2(\varphi,\lambda)\rbrack
\cr
+7\partial{c}c\partial\xi\xi{e^\phi}(2\partial{F_1}
(X,\psi)F_2(\varphi,\lambda)
-F_1(X,\psi)\partial{F_2}(\varphi,\lambda))\rbrace}}
for $k=3$.
Given the expressions (61)-(63), we are now prepared to analyze
the symmetry algebra generated
by ${\tilde{L}}^{(k)}$  that originate from $R_{2k}$
global $\alpha$-symmetries.
Just as 
the simplest global $\alpha$-generator ${\tilde{L}}_{\alpha+}(w)$
gives rise to local gauge symmetries, defined by
the BRST-exact derivatives $L_{\alpha+}^1=\partial{{\tilde{L}}_{\alpha+}}$ and 
$L_{\alpha+}^2=\partial^2{{\tilde{L}}_{\alpha+}}$,
the remaining generators of $R_2$, as well as those of $R_4$ and $R_6$
also give rise to their associate gauge symmetries.
For the remaining generators of $R_2$, ${\tilde{L}}_{\alpha\pm}$
and ${\tilde{L}}_{\alpha{m}}$,  structure of associate
gauge symmetries is simple
and similar to that of ${\tilde{L}}_\alpha$; the generators
are given by  $L_1^{(1)}\equiv\partial{\tilde{L}}^{(1)},
L_2^{(2)}\equiv\partial^2{\tilde{L}}^{(1)}$
with ${\tilde{L}}^{(1)}\equiv({\tilde{L}}_{\alpha\pm},{\tilde{L}}_{\alpha{m}})$
All the  gauge symmetry generators commute with each other
(just as in the case of a single ${\tilde{L}}_{\alpha+}$ generating
commutative ring with two elements
$L_{1,2}^\alpha$ considered above).
The $\alpha$-generators ${\tilde{L}}^{(2),(3)}$ of higher
cohomologies $R_4$ and $R_6$ give rise to the 
 local gauge symmetries with far more interesting structure.
Each of the $d+3$ elements of ${\tilde{L}}^{(2)}(w)\equiv
({\tilde{L}}_{\beta\pm}(w),{\tilde{L}}_{\beta\alpha}(w),
{\tilde{L}}_{\beta{m}}(w))$ gives rise to 4 
BRST exact gauge symmetry generators 
\eqn\grav{\eqalign{
L_k^{(2)}\equiv\partial^{k}{\tilde{L}}^{(2)}(w)=
\lbrace{Q_0},b_{-1}\partial^{k-1}{\tilde{L}}^{(2)}\rbrace\cr 
k=1,2,3,4}}
while each
 of the $d+4$ elements of ${\tilde{L}}^{(3)}(w)\equiv
({\tilde{L}}_{\gamma\pm}(w),{\tilde{L}}_{\gamma\beta},
{\tilde{L}}_{\gamma\alpha}(w),
{\tilde{L}}_{\gamma{m}}(w))$ gives rise to 6 
BRST exact gauge symmetry generators 
\eqn\grav{\eqalign{
L_k^{(3)}\equiv\partial^{k}{\tilde{L}}^{(3)}(w)=
\lbrace{Q_0},b_{-1}\partial^{k-1}{\tilde{L}}^{(3)}\rbrace\cr 
k=1,..,6}}
To determine the algebra of the gauge symmetry generators
${{L^{(n)}_{k}}}(n=1,2,3)$,
we first need to calculate the OPE's of their integrands.
We write
\eqn\lowen{{{L^{(n)}_{k}}}(w)=\partial_w^k\oint{{dz}\over{2i\pi}}
(z-w)^{2n}V^{(n)}(z)}
where the operators $V^{(n)}$ have conformal dimensions
$2n+1$ and
\eqn\grav{\eqalign{
V^{(1)}\equiv(V_{\alpha\pm},V_{\alpha{m}})\cr
V^{(2)}\equiv(V_{\beta\alpha},V_{\beta\pm},V_{\beta{m}})\cr
V^{(3)}\equiv(V_{\gamma\beta},V_{\gamma\alpha},V_{\gamma\pm},
V_{\gamma{m}})}}
so that $V_{\alpha\pm}$ is the integrand of ${\tilde{L}}_\pm$ etc.
As the manifest
expressions for $V^{(n)}$ operators (66), (67) are quite complicated,
we shall particularly concentrate on the subgroup of 9
$\alpha$-generators that are the Lorenz scalars, that is,
$({\tilde{L}}_{\alpha\pm},{\tilde{L}}_{\beta\pm},{\tilde{L}}_{\gamma\pm},
{\tilde{L}}_{\alpha\beta},{\tilde{L}}_{\alpha\gamma},{\tilde{L}}_{\beta\gamma})$.
The OPE calculation is quite cumbersome, although it can be somewhat
simplified by using the isomorphism between
the operators of $R_{2n}$ (with their truncated
versions being at minimal positive picture $n$)
 and those of the negative ghost cohomologies
$H_{-n-2}$, explained in ~{\selfgauge}.
The lengthy computation gives the following table of
the operator products:
\eqn\grav{\eqalign{
V_{\gamma\beta}(z_1)V_{\gamma\alpha}(z_2)
=...+{1\over{210}}\sum_{k=0}^6{{(-1)^k{k!}(6-k)!}\over{(6+k)!}}
{{\partial^{(k+2)}V_{\alpha\beta}(z_2)^{\lbrack{6}\rbrack}}\over
{(z_1-z_2)^{7-k}}}\cr
V_{\gamma\beta}(z_1)V_{\gamma\pm}(z_2)
=...+{1\over{210}}\sum_{k=0}^6{{(-1)^k{k!}(6-k)!}\over{(6+k)!}}
{{\partial^{(k+2)}V_{\beta\pm}(z_2)^{\lbrack{6}\rbrack}}\over
{(z_1-z_2)^{7-k}}}\cr
V_{\gamma\alpha}(z_1)V_{\gamma\pm}(z_2)
=...+{1\over{2520}}\sum_{k=0}^6{{(-1)^k{k!}(6-k)!}\over{(6+k)!}}
{{\partial^{(k+4)}V_{\alpha\pm}(z_2)^{\lbrack{6}\rbrack}}\over
{(z_1-z_2)^{7-k}}}\cr
V_{\gamma\beta}(z_1)V_{\beta\pm}(z_2)
=...+{1\over5}{{V^{\gamma\pm}}\over{(z_1-z_2)^5}}+
6\sum_{k=1}^4{{(-1)^k{k!}(4-k)!}\over{(6+k)!}}
{{\partial^{(k+2)}V_{\gamma\pm}(z_2)^{\lbrack{5}\rbrack}}\over
{(z_1-z_2)^{5-k}}}\cr
V_{\gamma\beta}(z_1)V_{\beta\alpha}(z_2)
=...+{1\over5}{{V^{\gamma\alpha}}\over{(z_1-z_2)^5}}+
6\sum_{k=1}^4{{(-1)^k{k!}(4-k)!}\over{(6+k)!}}
{{\partial^{(k+2)}V_{\gamma\alpha}(z_2)^{\lbrack{5}\rbrack}}\over
{(z_1-z_2)^{5-k}}}\cr
V_{\gamma\alpha}(z_1)V_{\beta\alpha}(z_2)
=...+{1\over5}{{V_{\gamma\beta}^{\lbrack{5}\rbrack}}\over{(z_1-z_2)^5}}+
6\sum_{k=1}^4{{(-1)^k{k!}(4-k)!}\over{(6+k)!}}
{{\partial^{(k+2)}V_{\gamma\beta}(z_2)^{\lbrack{5}\rbrack}}\over
{(z_1-z_2)^{5-k}}}\cr
V_{\gamma\pm}(z_1)V_{\beta\mp}(z_2)
=...+{1\over5}{{V_{\gamma\beta}^{\lbrack{5}\rbrack}}\over{(z_1-z_2)^5}}+
6\sum_{k=1}^4{{(-1)^k{k!}(4-k)!}\over{(6+k)!}}
{{\partial^{(k+2)}V_{\gamma\beta}(z_2)^{\lbrack{5}\rbrack}}\over
{(z_1-z_2)^{5-k}}}\cr
V_{\alpha\pm}(z_1)V_{\gamma\alpha}(z_2)
=...+{1\over3}{{V_{\gamma\pm}^{\lbrack{4}\rbrack}}\over{(z_1-z_2)^3}}-
{1\over{42}}{{\partial{V_{\gamma\pm}^{\lbrack{4}\rbrack}}}\over{(z_1-z_2)^2}}
+{{1}\over{168}}{{\partial^2{V_{\gamma\pm}^{\lbrack{4}\rbrack}}}\over
{z_1-z_2}}
\cr
V_{\alpha\pm}(z_1)V_{\gamma\pm}(z_2)
=...+{1\over3}{{V_{\gamma\alpha}}\over{(z_1-z_2)^3}}-
{1\over{42}}
{{\partial{V_{\gamma\alpha}^{\lbrack{4}\rbrack}}}\over{(z_1-z_2)^2}}
+{{1}\over{168}}{{\partial^2{V_{\gamma\alpha}^{\lbrack{4}\rbrack}}}\over
{z_1-z_2}}
\cr
V_{\beta\alpha}(z_1)V_{\beta\pm}(z_2)
=...+{1\over{60}}\sum_{k=0}^4{{(-1)^k{k!}(4-k)!}\over{(4+k)!}}
{{\partial^{(k+2)}V_{\alpha\pm}(z_2)^{\lbrack{4}\rbrack}}\over
{(z_1-z_2)^{5-k}}}
\cr
V_{\beta\alpha}(z_1)V_{\alpha\pm}(z_2)
=...+{1\over3}{{V_{\beta\pm}^{\lbrack{3}\rbrack}}\over{(z_1-z_2)^3}}-
{1\over{30}}
{{\partial{V_{\beta\pm}^{\lbrack{3}\rbrack}}}\over{(z_1-z_2)^2}}
+{{1}\over{90}}{{\partial^2{V_{\beta\pm}^{\lbrack{3}\rbrack}}}\over
{z_1-z_2}}
\cr
V_{\beta\pm}(z_1)V_{\alpha\mp}(z_2)
=...+{1\over3}{{V_{\beta\alpha}^{\lbrack{3}\rbrack}}\over{(z_1-z_2)^3}}-
{1\over{30}}
{{\partial{V_{\beta\alpha}^{\lbrack{3}\rbrack}}}\over{(z_1-z_2)^2}}
+{{1}\over{90}}{{\partial^2{V_{\beta\alpha}^{\lbrack{3}\rbrack}}}\over
{z_1-z_2}}
}}
where the numbers in the square brackets on the right hand side
refer to the superconformal pictures of the generators and
we have skipped  the OPE terms that are too singular to
contribute to the algebra of the $\alpha$-generators ${\tilde{L}}^{ij}(w)$
(where the $i,j$ indices stand for $\alpha,\beta$ or $\gamma$)
That is, for example, the commutator of
${\tilde{L}}_{\gamma\beta}(w)=\oint{{dz_1}\over{2i\pi}}
(z_1-w)^6{V_{\gamma\beta}}(z_1)$ with  
${\tilde{L}}_{\gamma\alpha}(w)=\oint{{dz_2}\over{2i\pi}}
(z_2-w)^6{V_{\gamma\alpha}}(z_2)$ won't be contributed
by the terms in the OPE of
${V_{\gamma\beta}}(z_1)$ and ${V_{\gamma\alpha}}(z_2)$
 with the singularity order of $(z_1-z_2)^{-8}$ or stronger,
therefore we  skip these terms in the first OPE in the table (68),
starting the expansion from the order of $(z_1-z_2)^{-7}$.
Using the OPE table (68) it is not difficult to compute the algebra
of the global $\alpha$-generators and of the local gauge symmetry
generators $L_{ij}^m\equiv\partial^m_w{\tilde{L}}_{ij}$
of the associate gauge symmetries (as before, the $m$ index
runs from 1 to $2k$ for the gauge symmetries associated with the
generators of $R_{2k}$; that is, $k=1$ for $L_{\alpha\pm}^m$,
$k=2$ for $L_{\beta\pm}^m,L_{\beta\alpha}^m$ and $k=3$
for $L_{\gamma\pm}^m,L_{\gamma\alpha}^m,L_{\gamma\beta}^m$.
First of all, simple calculation using the OPE's (68) shows that the
9 global $\alpha$-generators,
${\tilde{L}}_{ij}\equiv({\tilde{L}}_{\alpha\pm},{\tilde{L}}_{\beta\pm},
{\tilde{L}}_{\gamma\pm},
{\tilde{L}}_{\beta\alpha},{\tilde{L}}_{\gamma\alpha},
{\tilde{L}}_{\gamma\beta})$ satisfy
the commutation relations of $U(3)$:
\eqn\lowen{{\lbrack}{\tilde{L}}_{i_1j_1},{\tilde{L}}_{i_2j_2}\rbrack=
-\delta_{i_1j_2}{\tilde{L}}_{i_2j_1}+\delta_{i_2j_1}{\tilde{L}}_{i_1j_2}}
The computation of commutators of the $\alpha$-generator's derivaves
$L_{ij}^m$ using the table (68) is straightforward as well.
It is convenient to redefine the generators $L_{ij}^m\rightarrow{T_{ij}^m}$
according to
\eqn\grav{\eqalign{
T_{\alpha\pm}^m={{L_{\alpha\pm}^m}\over{(2-m)!}}(m=1,2)\cr
T_{\beta\alpha}^m={{L_{\beta\alpha}^m}\over{(4-m)!}}(m=1,2,3,4)\cr
T_{\beta\pm}^m={{L_{\beta\pm}^m}\over{(4-m)!}}(m=1,2,3,4)\cr
T_{\gamma\alpha}^m={{L_{\gamma\alpha}^m}\over{(6-m)!}}(m=1,...,6)\cr
T_{\gamma\beta}^m={{L_{\gamma\beta}^m}\over{(6-m)!}}(m=1,...,6)\cr
T_{\gamma\pm}^m={{L_{\gamma\pm}^m}\over{(6-m)!}}(m=1,...,6)}}
Then the commutators of $T_{ij}^m$ satisfy
\eqn\grav{\eqalign
{{\lbrack}T_{i_1j_1}^m,T_{i_2j_2}^n\rbrack=
(n-m)(\delta_{i_2j_1}T_{i_1j_2}^{m+n}-\delta_{i_1j_2}T_{i_2j_1}^{m+n})}}
provided that  of $m+n\leq{2k}$
where $2k_{r.h.s.}$ is the order of the $R_{2k_{r.h.s.}}$ cohomology of each
corresponding operator on the right hand side;
Otherwise, in case if $m+n>2k_{r.h.s.}$, the generators commute.
Schematically,
\eqn\grav{\eqalign{{\lbrack}T_I^m,T_J^n\rbrack
=(m-n){f_{IJ}^{K}}T_K^{m+n} (m+n\leq{2k_{r.h.s.}})\cr
 (m+n\leq{2k_{r.h.s.}})  }}
where the capital indices stand for $I=(i_1,j_1), J=(i_2,j_2)$, etc;
$f_{IJ}^K$  are the $U(3)$ structure constants, so the 
algebra of the gauge symmetries  associated to
$R_{2k} (k=1,2,3)$ $\alpha$-generators is isomorphic to $U(3)\times{X_6}$
where $X_6$ is solvable Lie algebra consisting
of 6 elements $x_1,...,x_6$  with the commutation
relations given by
\eqn\grav{\eqalign{\lbrack{x_m},{x_n}\rbrack=(m-n)x_{n+m}(m,n=1...6;m+n\leq{6})
\cr
{\lbrack{x_m},{x_n}\rbrack=0 ( m+n>6)}}}
Given the $T_{ij}^m$ generators
of the $U(3)\times{X_6}$ gauge symmetries
, it isn't difficult to show that
the generalized $b$-ghost fields (in the adjoint of
$U(3)\times{X_6}$) satisfying
\eqn\lowen{L_{ij}^m=\lbrace{Q_0},\partial^{m-1}\oint{B_{ij}}\rbrace}
 are given by
\eqn\grav{\eqalign{
{\oint}B_{ij}=b_{-1}{\tilde{L}}_{ij}}}
similarly to the simplest case (35).
Given the generalized $b$-ghost fields of (74), one  can construct
 the generalized $c$-ghost fields
$C_{ij}$, conformal dimension $-1$ primaries and canonical
conjugates of $B_{ij}$, satisfying
\eqn\lowen{\lbrace\oint{B_{ij}},C_{ij}\rbrace=:\Gamma^n:}
where $\Gamma$ is again the picture-changing operator
with $n=1,2,3$ for ${\tilde{L}}_{ij}$'s of $R_2$, $R_4$
and $R_6$ respectively.
With some effort, it is possible to derive explicit expressions
for 
 various $C_{ij}$'s. For the $C_{ij}$'s corresponding
to ${\tilde{L}}_{ij}$'s of $R_4$ 
the expressions are given by

\eqn\grav{\eqalign{C_{ij}^{(R_4)}
=e^\phi{G}e^{3\phi-\chi}L^{matter}_{ij({9\over2})}
P^{(1)}_{\phi-{{10}\over3}\chi}
+\lbrace{{\tilde{Q}}_0},\partial^2{b}\partial{b}b{e^{5\phi-2\chi}}
{L^{matter}_{ij({9\over2})}}
\rbrace}}
where 
$F^{matter}_{ij(5)}$ are the
conformal dimension 5  matter parts of the ${\tilde{L}}_{ij}$
operators of $R_4$ given in (51),
$L^{matter}_{ij({9\over2})}$ are the worldsheet superpartners of
$F^{matter}_{ij(5)}$ satisfying
$$\oint{{dz\over{2i\pi}}}
\lbrace{G}(z),L^{matter}_{ij({9\over2})}(w)\rbrace=F^{matter}_{ij(5)}(w)$$
and ${\tilde{Q}}_0$ is the ghost part of the standard BRST charge:
\eqn\lowen{
{\tilde{Q}}_0=Q_0-\oint{{dz}\over{2i\pi}}(cT-bc\partial{c})}
The generalized $C_{ij}$-ghost operators (77) associated with the 
  $R_4$ gauge symmetry generators (70) are the ``elements
of $H_3$'' (off-shell operators existing at
superconformal ghost pictures 3 and above, but not below 3)
satisfying the canonical relations
with the $R_4$-associated $B_{ij}$-ghosts, up to the double picture-changing:
\eqn\lowen{\lbrace\oint{{dz}\over{2i\pi}}B_{ij{\lbrack}-1{\rbrack}}^{(R_4)},
C_{ij}^{(R_4)}\rbrace=:\Gamma^2:}
Next, the expressions the generalized $c$-ghost fields $C_{ij}$,
corresponding to the gauge symmetries 
associated with the elements (70) of $R_6$ are
given by
\eqn\grav{\eqalign{C_{ij}^{(R_6)}=
:e^\phi{G}e^{4\phi-\chi}P^{(1)}_{\phi-3\chi}L_{ij(9)}^{matter}:
\cr
+\lbrace{\tilde{Q}}_0,\partial^4{b}\partial^3{b}\partial^2{b}\partial{b}{b}
e^{7\phi-3\chi}(\partial{P^{(1)}_{\phi-{9\over4}\chi}}
+
{P^{(1)}_{\phi-{9\over4}\chi}}P^{(1)}_{7\phi-3\chi-7\sigma})
F_{ij({{17}\over2})}^{matter}\rbrace}}
where the conformal dimension ${{17}\over2}$
primaries
$F_{ij({{17}\over2})}^{matter}$ are the matter components of
the $R_6$-generators of (54), (61), (63) while the conformal dimension $9$
primaries $L_{ij(9)}^{matter}$ are the worldsheet superpartners
of 
$F_{ij({{17}\over2})}^{matter}$, satisfying
$$\lbrace\oint{{dz\over{2i\pi}}}
{G}(z),F^{matter}_{ij({{17}\over2})}(w)\rbrace=L^{matter}_{ij(9)}(w).$$
The generalized $C_{ij}$-ghost operators (80) associated with the 
derivatives  of the  $R_6$ $\alpha$-generators (54),(61),(63) are 
the ``elements
of $H_4$'' (off-shell operators existing at
superconformal ghost numbers 4 and above, but not below 4)
satisfying the canonical relations
with the $R_6$-associated $B_{ij}$-ghosts, up to the 
triple picture-changing:
\eqn\lowen{\lbrace\oint{{dz}\over{2i\pi}}B_{ij{\lbrack}-1{\rbrack}}^{(R_6)},
C_{ij}^{(R_6)}\rbrace=:\Gamma^3:}
Finally, given the $U(3)\times{X_6}$ gauge symmetries 
 induced by  $T^m_I\equiv{T^m_{ij}}$ (capital indices stand for the abbreviation
of (ij), as previously),
 the generalized ghosts $B_I\equiv{B_{ij}}$,
$C_I{\equiv}C_{ij}$ and the symmetry algebra (71), (72)
 (70), (75), (77) it is straightforward to construct the nilpotent BRST
operator associated with the 
U(3) subgroup of $\alpha$-symmetries of $R_{2n} (n=1,2,3)$:
\eqn\grav{\eqalign{Q_2=\sum_{I}
{\sum_{m=1}^6}(C_I^{})_mT^m_I+{1\over2}
\sum_{I,J,K}\sum_{m,n=1}^{m+n\leq{6}}(m-n)f^{IJK}({C_I^{}})_m
({C_J^{}})_n({B_K^{}})_{-m-n}}}
where $f^{IJK}$ are again the appropriate
$U(3)$ structure constants and

\eqn\grav{\eqalign{C_I^n=\oint{{dz}\over{2i\pi}}
z^{m-2}C_I(z)\cr
(B_I)_m=m!\oint{{dw}\over{2i\pi}}{{b_{-1}{\tilde{L}}_K(w)}\over{{w^{m+1}}}}}}

Depending on the values of $I,J$ and $K$  of the $U(3)$ indices in
(82),
the generalized $B_I$ and $C_I$ ghosts entering the expression
for $Q_2$ are related to the various classes of the gauge symmetries
derived from the $R_2$, $R_4$ or $R_6$ cohomologies
(with the precise expressions given in (43),(44),(74), (77)
and (80) for each cohomology case).

While the manifest integral form expression for $Q_2$
is quite complicated (far more complicated and lengthy than
for $Q_1$ of $H_1$ constructed in (48)), it can be shown
that structurally it consists of three terms which are
on-shell (with respect to $Q_0$) and are the elements
of superconformal ghost cohomologies $H_1$, $H_2$ and $H_3$
respectively; while the $H_1$ part of $Q_2$ 
contains $Q_1$ of (48), the $H_2$ and $H_3$ ingrediants
are related to the gauge symmetries associated with the geometry
$\alpha$-generators of $R_4$ and $R_6$.
The BRST operator $Q_2$ (82) 
can be shown to commute with 
the nilpotent BRST charges $Q_0$ and $Q_1$ 
of the lower ghost cohomologies. It appears that $Q_2$ describes 
RNS strings 
in curved background which is shaped by the geometry
of the extra dimensions induced by the $U(3)$ subgroup of the 
$\alpha$-generators. We leave the detailed analysis
of this geometry, as well as the analysis of the cohomologies
of $Q_1$ and $Q_2$, for the future work.
It would be very interesting to
 generalise the construction described in this paper in order to
 build the nilpotent BRST charges based on the
higher order $b-c$ ghost cohomologies $R_{2n}(n\geq{4})$.
The expressions for the $\alpha$-generators involving
the higher ghost numbers are, however, increasingly complicated;
understanding their structures clearly requires further effort.

\centerline{\bf 6. Conclusion and Discussion}

In this paper we constructed a sequence
of new nilpotent BRST charges $Q_1$ (48) and $Q_2$ (82) 
consisting of the currents in superconformal ghost cohomologies
$H_1$, $H_2$ and $H_3$. The construction is based on the sequence
of hidden local gauge symmetries of the RNS theory,
associated with the ground ring of $\alpha$-generators,
that are classified in terms of the $b-c$ ghost cohomologies
$R_2$, $R_4$ and $R_6$. The  $\alpha$-generators,
 in turn, induce global space-time symmetries originating
from hidden  space-time dimensions.
The constructed BRST charges thus appear to describe
superstring theories in various curved backgrounds
which geometry is shaped by the extra dimensions
induced by the global $\alpha$-symmetries
in $R_{2n} (n=1,2,3)$. 
In principle, the construction of the nilpotent BRST charges
charges could be extended to higher $n>3$, as there is no clear 
reason to exclude the existence of $\alpha$-symmetry generators and 
their ground rings at $b-c$ cohomology levels  $R_{2n}$ with $n>3$.
At present, however, we do not have the construction for the $n>3$
cases, as the expressions for the generators and their OPE's
become extremely complicated at $n>3$ cases, with the technical
difficulties always aggravated by the picture-changing.
Our preliminary analysis ~{\progress}
shows, however, that there exists certain underlying
principle defining the structures 
of the $\alpha$-generators of higher cohomologies;
if one is able to advance in this direction, this could probably
simplify the construction of the BRST charges for higher $n$'s.
As was noted above,
the new BRST charges constructed in this paper define the sequence of RNS
string theories constructed in certain curved geometrical backgrounds,
typically $AdS$ (in the case of $Q_0+Q_1$)
 or $AdS\times{CP_n}$-type (in the case of 
linear combination of $Q_0$, $Q_1$ and $Q_2$)~{\progress};
these backgrounds have to be identified by the constraints that the gauge
transformations, induced by the ground ring of the $\alpha$-symmetries,
impose on space-time geometry. 
The work concerning this problem is currently
in progress and we hope to clarify the related questions in future works.
Since the charges $Q_0$, $Q_1$ and $Q_2$ commute with each other, their
combinations , such as $Q_0+Q_1$, $Q_0+Q_2$ 
or $Q_0+Q_1+Q_2$ define kinetic terms
in certain RNS string field theories (SFT)
built around the curved backgrounds, mentioned above.
For example, the preliminary analysis of the
cohomology of $Q_0+Q_1$ charge in open string theory
 shows that its cohomology consists of a single gauge boson:
\eqn\grav{\eqalign{
V(k)=\oint{e^{-3\phi}}\lbrace
({\vec{A}}\partial{\vec{X}})
({\vec{k}}\partial{\vec{X}})
({\vec{k}}{\vec{\psi}})
+
({\vec{A}}{\vec{\psi}})
({\vec{k}}\partial{\vec{\psi}})
({\vec{k}}{\vec{\psi}})
({\vec{A}}{\vec{\psi}})
({\vec{k}}\partial{\vec{X}})^2{\rbrace}{e^{i{\vec{k}}{\vec{X}}}}
\cr
{\vec{k}}{\vec{A}}({\vec{k}})=0; ({\vec{k}})^2=0}}
with no massive states at all. Such a spectrum
is typical for strings in $AdS$-type background
which are known to be dual to gauge theories ~{\malda, \wit, \ampf, \ampconf}.
In general, we 
 expect the backgrounds,
 corresponding to new BRST charges found in this paper,
to include those of the $AdS$ or $AdS_q\times{CP_n}$-type
~{\progress},  with the $CP_n$-structures related
to subgroups of the $\alpha$-symmetries restricted to
certain ghost cohomology classes.
The particular background geometries, however,
 depend on  various choices
of different subgroups or cosets of the underlying $\alpha$-symmetries.
For example, the $CP_2\sim{{SU(3)}\over{{SU(2)\times{U(1)}}}}$
fibration can be obtained  by taking the $U(3)$-subgroup (67)
of the $\alpha$-generators, factorizing
by $SU(2)$ generated by ${\tilde{L}}_{\alpha\beta},
{\tilde{L}}_{\alpha\gamma}$ and ${\tilde{L}}_{\alpha\beta}$,
excluding ${\tilde{L}}_{\alpha+}$ (playing the role of $U(1)$)
and defining certain linear combinations
of the remaining $\alpha$-transformations up
to $L_{\alpha-}$ (which is possible since the latter commutes
with physical vertex operators ~{\selfgauge})

Exploring the new SFT's based on these
new BRST charges could be  useful in order to develop
the non-supersymmetric versions of AdS/CFT. In particular,
in the $d=4$ case the gauge-string duality can be understood
as the duality between closed SFT built 
around the $AdS_5$ background and the loop
equations in $d=4$
 ~{\slava} , with the BRST charges of the curved string field
theory 
being dual to the loop operator in non-supersymmetric QFT.
On the other hand, studying the string field theories with
BRST charges related to
 $AdS_q\times{CP}_n$ structures would be useful to explore
some other remarkable examples of $AdS/CFT$ dualities,
such as  the duality type 
between IIA strings on $AdS_4\times{CP_3}$  and
the 't Hooft  limit of three-dimensional $SU(N)\times{SU}(N)$
gauge theory describing the effective worldvolume dynamics 
of M2 branes ~{\bagger, \abjm}.
 In particular, it would be interesting to relate
the symmetries of Bagger-Lambert and ABJM three-dimensional 
theories to the gauge symmetries associated
to  appropriate ground rings of $\alpha$-generators. This,
however, would require a better understanding of the $R_8$
cohomology of operators which structure is still obscure.
Another challenging development would be to construct BRST
charges that would imitate $M$-theory dynamics on $AdS_7\times{CP_2}$
that could perhaps shed some light on the $M5$-brane worldvolume physics. 
We hope that the results presented in this paper will be helpful
for the progress in these directions.

\centerline{\bf Acknowledgements}

I would like to thank  Robert De Mello Koch, Kevin Goldstein,
Antal Jevicki,
Sanjaye Ramgoolam and
 Joao Rodrigues for interesting productive discussions.
I also would like to thank Kevin Goldstein and Anthony Germishuys 
for their help in recovering the text of this paper after
it had been nearly lost due to major Linux crash on my computer.

\listrefs

\end